\documentclass{article}

\usepackage{PRIMEarxiv}

\usepackage[utf8]{inputenc} 
\usepackage[T1]{fontenc}    
\usepackage{hyperref}       
\hypersetup{
    colorlinks=true,
    linkcolor=blue,
    filecolor=magenta,      
    urlcolor=cyan,
}
\usepackage{url}            
\usepackage{doi}
\usepackage{booktabs}       
\usepackage{amsfonts}       
\usepackage{nicefrac}       
\usepackage{microtype}      
\usepackage{lipsum}
\usepackage{fancyhdr}       
\usepackage{graphicx}       
\graphicspath{{media/}}     

\usepackage{multirow}
\usepackage{textcomp,mathcomp}
\usepackage{float}
\usepackage{lscape}

\title{Personalized local heating neutralizing individual, spatial and temporal thermo-physiological variances in extreme cold environments
\thanks{Manuscript submitted to \emph{Building and Environment} on 10 Oct. 2022. Accepted 19 Dec. 2022.\\
\indent \indent DOI: \url{https://doi.org/10.1016/j.buildenv.2022.109950}. \emph{updated on 27 Dec. 2022.}}
}

\usepackage{authblk}

\author[1,2, $\dagger$]{Yi Ju}
\author[1,3, $\dagger$]{Xinyuan Ju}
\author[2]{Hui Zhang}
\author[1,3, *]{Bin Cao}
\author[4]{Bin Liu}
\author[1,3]{Yingxin Zhu}

\affil[1]{\footnotesize Department of Building Science, School of Architecture, Tsinghua University, Beijing 100084, China
}
\affil[2]{\footnotesize Center for the Built Environment (CBE), University of California, Berkeley, CA, 94720, USA 
}
\affil[3]{\footnotesize Key Laboratory of Eco Planning and Green Building, Ministry of Education (Tsinghua University), Beijing, 100084, China}
\affil[4]{\footnotesize Key Laboratory of Refrigeration Technology, Tianjin University of Commerce, Tianjin, 300134, China
}
\affil[*]{Corresponding author: Bin Cao, \tt caobin@tsinghua.edu.cn}
\affil[$\dagger$]{These two authors contribute equally to this paper}

\usepackage{amsmath}
\allowdisplaybreaks

\usepackage{setspace}

\usepackage{lineno}

\begin{document}
\maketitle

\allowdisplaybreaks

\begin{abstract}
In this paper, we investigate the feasibility, robustness and optimization of introducing personal comfort systems (PCS), apparatuses that promises in energy saving and comfort improvement, into a broader range of environments. We report a series of laboratory experiments systematically examining the effect of personalized heating in neutralizing individual, spatial and temporal variations of thermal demands. The experiments were conducted in an artificial climate chamber at -15 $\tccentigrade$ in order to simulate extreme cold environments. We developed a heating garment with 20 pieces of 20 × 20 cm² heating cloth (grouped into 9 regions) comprehensively covering human body. Surface temperatures of the garment can be controlled independently, quickly (within 20 seconds), precisely (within 1 $\tccentigrade$) and easily (through a tablet) up to 45 $\tccentigrade$. Participants were instructed to adjust surface temperatures of each segment to their preferences, with their physiological, psychological and adjustment data collected. We found that active heating could significantly and stably improve thermal satisfaction. The overall TSV and TCV were improved 1.50 and 1.53 during the self-adjustment phase. Preferred heating surface temperatures for different segments varied widely. Further, even for the same segment, individual differences among participants were considerable. Such variances were observed through local heating powers, while unnoticeable among thermal perception votes. In other words, all these various differences could be neutralized given the flexibility in personalized adjustments. Our research reaffirms the paradigm of “adaptive thermal comfort” and will promote innovations on human-centric design for more efficient PCSs.
\end{abstract}


\section{Introduction}
\subsection{Background}

Thermal comfort research in extreme cold environments is limited. Widely-used indices such as PMV, SET* are not applicable to evaluate people’s thermal sensation in such environments \cite{chen_investigation_2020}, \cite{du_field_2020}. Meanwhile, research on applications to improve thermal comfort is greatly in need, since there are numerous people facing extreme cold exposure, such as cold storage workers, border guards, ice sports spectators, etc.

Lessons from the extensively studied indoor thermal comfort over the past few decades remind us of the common dilemma of energy consumption and comfort. The temperature in most buildings is controlled in a narrow range by indoor space conditioning to meet the thermal comfort criterion based on predicted mean vote (PMV) proposed in the 1970s \cite{fanger_thermal_1970}, which incurs a significant energy cost \cite{brager_evolving_2015}\cite{zhang_using_2015}. Moreover, research shows that even within the narrowly-controlled temperature range, thermal dissatisfaction is still a major concern \cite{arens_are_2010},\cite{graham_lessons_2021}, given the considerable diversity of individual thermal sensitivities and preferences \cite{wang_individual_2018}, \cite{luo_high-density_2020}. 

Compared to conventional thermal environment control methods, personal comfort systems (PCS) can satisfy personalized demand with energy-saving potential \cite{yang_thermal_2022} because they only condition the micro-environment surrounding target occupants instead of the whole room space. Besides, since portable PCS can maintain comfort locally, it is promising to extend applications from indoors to outdoors. 

Multiple thermo-physiological and epidemiological studies confirm that cold exposure may bring harmful impacts on people's health. Nasal discharge, sneezing, shivering, skin pain and numbness are common symptoms during cold winter \cite{chen_physiological_2019}, \cite{saedpanah_effects_2019}, and longer exposure may result in major injuries to hand and finger flexibility and sensorineural functioning \cite{wu_human_2021}, \cite{thetkathuek_cold_2015}. Cold temperature and sudden temperature drop also plays a risky role of asthma exacerbation and some other respiratory symptoms \cite{zhu_cold_2022}. In cold conditions, the cardiovascular system is under considerable strain since the body will experience a series of physiological reactions such as vasoconstriction and decreased blood flow after the cold signal is perceived by the posterior hypothalamic cold receptors before cold shivering occurs \cite{tipton_human_2017}. Experiments find that people exposed to extreme cold environments (-20$\tccentigrade$) could result in a 17.6\% blood pressure increase within a minute \cite{wu_physiological_2021}. Thus, the incidence rate of ischemic stroke \cite{chen_underlying_2022}, myocardial infarction \cite{cai_cold_2016}, high blood pressure\cite{wu_physiological_2021-1}  etc., are significantly increased under extreme cold conditions. What’s more, there also exists research finding indicating an inverse relationship between ambient temperature and cancer risk \cite{bandyopadhayaya_can_2020}. In addition to physical influence, tension, anger, confusion, decreased depression and fatigue will also increase during exposure to extremely cold environments \cite{wu_perceptual_2021}.

Though people can acclimate to cold after prolonged exposure \cite{lichtenbelt_cold_2014}, it should be highlighted that humans' physiological ability to resist the negative health impacts of cold is limited, even among people who have acclimated to the cold climate \cite{saedpanah_effects_2019}. Therefore, measures like meteorological early warning \cite{chen_underlying_2022}, establishing transition space \cite{wu_physiological_2021-1}, and personal comfort systems (PCS) are proposed to relieve the health risk caused by cold exposure \cite{luo_effects_2022}. Compared to other methods, PCS shows irreplaceable characters that enable people to meet their various personalized thermal needs, even without the shelter of building envelopes or conventional HVAC systems. Furthermore, different regions of the human body contribute differently to thermal comfort \cite{nakamura_relative_2013}, \cite{zolfaghari_thermal_2010} and there are physiological interactions among the body parts, for example, when one region of the body cools down, it triggers reflex vasoconstriction in other regions \cite{castellani_human_2016}. Therefore, research exploring simultaneous thermal stimulus across human body is more expected rather than measuring thermal reactions on an isolated zone.

\subsection{Related work}

Above provides a brief introduction on PCS’s fundamentals and applications, the related thermo-physiological and psychological research, as well as some specific research in cold environments. In this section, we are to further review some of the most related works to our research. Based on the review, we find the existence of three main research gaps in current research that our research would make contributions to.

To perform the review in this section, we searched literature published in the last ten years on SCI-sourced journals with a reasonable number of citations. We selected laboratory experiment research on personalized heating / cooling which has both physiological measurements and subjective questionnaires. A total of 26 literatures are included, and are evaluated based on the following three questions:

\begin{itemize}
    \item Which body parts are locally heated / cooled?
    \item What kinds of local (heating) combinations have been experimented?
    \item Can personal / local heating be controlled by participants?
\end{itemize}

We found that the scope and depth of PCS research in recent years experienced great improvement. They include but are not limited to: Thorough and detailed physiological experiments were conducted to understand local thermal sensitivity distributions, e.g., \cite{luo_high-density_2020}. Local heating / cooling combinations were explored, e.g., \cite{deng_effects_2019}, \cite{wang_experimental_2020}, \cite{he_creating_2022}. Researchers investigated how personal control by users can potentially meet their diverse needs and even create thermal pleasures, e.g., \cite{he_creating_2022}, \cite{he_meeting_2021}.

Meanwhile, we also define the following three research gaps to bridge.

\subsubsection*{Thermal comfort characteristics at segment level have not been fully investigated.}

The success of PCS comes from the recognition that people’s thermal demands are diverse and they can be satisfied on individual scales. There has been plenty of evidence suggesting that thermal demands of different body parts also vary a lot \cite{luo_high-density_2020}, \cite{luo_effectiveness_2022}. However, investigations on the characteristics of different body parts from thermal demand-response perspective are not comprehensive yet.

Due to the diversity of PCS formats and products, results among different experiments are also largely incomparable. Most previous research suggests some particular type of PCS works well \cite{luo_thermal_2018}, \cite{tang_thermal_2022}, however, it’s time to think of the question in a more scientific manner and answer it more explicitly: what level of local heating / cooling is needed by a particular body part under some given ambient environment - which can be produced by multiple kinds of devices, or even devices we haven’t seen.

\subsubsection*{There lacks a systematic and efficient method for evaluation and optimization of local heating / cooling combinations.}

Yang et al’s  \cite{yang_study_2020} experiment shows that the comfort improvement is limited when only applying local cooling on chest or buttocks area solely, but can be further improved when considering the combination of the two body parts. The research also concludes that there usually exists a tradeoff between improvement in overall thermal sensation and the local overcooling discomfort when local cooling is only applied on one or two body parts. Luo et al’s \cite{luo_thermal_2018} research shows that a proper combination of several low-power portable comfort devices can perform as powerful as central air conditioning. Some researchers were also motivated to find the optimal combinations of local heating / cooling in terms of e.g., comfort, devices’ total power, weight and capital cost \cite{tang_thermal_2022}, but no systematic approach has been proposed.
 
A fundamental challenge is the so-called “curse of dimensionality” that a comprehensive comparison requires an exponentially increasing number of experiments with the number of body parts investigated. For instance, Wang et al \cite{wang_experimental_2020} tested the performance of local heating devices applied on three body parts, where in order to compare all the combinations, eight groups of experiments were conducted. If we are to consider 10 body parts / devices, theoretically a total of 1024 on/off combinations are available, letting alone adjustments on power.

\subsubsection*{Flexibility of personalized control by participants can be further enhanced and utilized.}

Although more and more recent research allows and/or even encourages participants to make some adjustments on the comfort devices, participants’ interaction is still very limited.

First, it’s not surprising that some of the most comprehensive experiments, e.g., Zhang et al’s \cite{zhang_thermal_2010-2}\cite{zhang_thermal_2010-1}\cite{zhang_thermal_2010}\cite{zhao_thermal_2014} in developing the local comfort model, Luo et al’s \cite{luo_high-density_2020} on local thermal sensitivity, do not consider personal adjustments at all in order to make their experiments well under control. 

Second, most adjustment options found in literature are at the device level, i.e., users can only adjust the total power of devices \cite{pasut_energy-efficient_2015}, \cite{rissetto_personalized_2021} but have no access to control local heating / cooling power on particular body segments. Further, among those allowing personal adjustments, most just average participants’ choices while individual differences in heating / cooling demands have not been carefully analyzed \cite{arghand_individually_2022}, \cite{he_meeting_2021}. Understanding individual differences and then providing adequate choices for users is crucial to make sure that PCS can meet the demands of every occupant.

Thirdly, there has been a rich collection of literature investigating the psychological effects of personal control on thermal comfort \cite{luo_can_2014}. Concepts such as “perceived control” \cite{luo_underlying_2016}, “alliesthesia” \cite{parkinson_predicting_2021} have been proposed, which are not only related to comfort, but also pleasure. However, still few studies formally introduce these effects into their design, behind which is the old-fashioned paradigm that treats people as passive subjects instead of active participants \cite{brager_thermal_1998}.

To conclude, we are interested in understanding: first, which body parts are of the most need of local heating in cold environments; second, what is the thermal demand distribution like across all body parts; third, how are individual differences on the segment level, and can personalized adjustment make PCS satisfy everyone’s need. Our study is aimed at tackling aforementioned challenges systemically with a novel approach of experimental design and system.

\subsection{Objectives / contributions}

Our research aims at exploring the thermal demand distributions of the human body in cold environments, with an emphasis on the individual, spatial and temporal thermo-physiological variances. In order to evaluate and satisfy such dynamic and diverse thermal requirements, we introduce a new experiment design which enables participants to have full control on the localized heating elements. Our research, which will enlighten applications on wearable PCS design in cold environments, also contributes as fundamental research to human thermal comfort which will hopefully refresh research paradigms of the community. Our main contributions lie in following aspects:

\begin{itemize}
    \item We developed a novel system for local comfort research, including modularized heating fabrics with high efficiency and accuracy, together with portable user-interface from which participants can make instant and independent power adjustment on each of the nine body heating areas. We proposed a fully participant-centric experiment design based on our system.
    \item We investigated local thermal demand distributions of the human body by allowing any possible combinations of different heating power on different body segments, reducing the inductive bias in experiment design.
    \item We systematically evaluated the individual, spatial and temporal thermo-physiological variances, and provided evidence that personalized adjustments satisfy such dynamic and diverse thermal requirements and even create thermal pleasure. We proposed the “adaptive control” framework to summarize how personalized and localized control work.
\end{itemize}

The following parts of this paper will be organized as follows: In section 2, we will introduce the experiment design and basic information (chamber, participants, devices), together with a thorough introduction to the intelligent personal comfort system. In section 3, we report the results of our experiments, including detailed analysis on participants’ adjustment behaviors, physiological responses and subjective evaluations. In section 4, we compare our results with previous findings, propose an explanatory framework known as “adaptive adjustment”, and discuss the limitations and future work. Section 5 is the conclusion section of the paper.

\section{Method} \label{sec:method}
\subsection{Basic experiment information}

\subsubsection*{Experiment setup}

\begin{figure}[h]
    \centering
    \includegraphics[width=16cm]{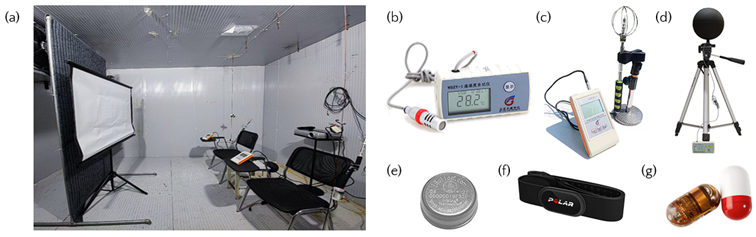}
    \caption{Climate chamber, environment measurement devices and physiological measurement devices. \small \textbf{a}. Layout of the artificial climate chamber. \textbf{b}. self-recording thermometer and hygrometer. \textbf{c}. anemometer. \textbf{d}. black bulb thermometer. \textbf{e}. skin temperature sensor. \textbf{f}. heart rate sensor. \textbf{g}. core body temperature capsule.}
    \label{fig:fig1}
\end{figure}

The experiments were conducted in an artificial climate chamber in Tianjin University of Commerce, China, which is particularly designed to simulate cryogenic environments. The chamber is 3.62 m × 2.67 m in area and 2.29 m in height (Figure \ref{fig:fig1}.a). The lowest temperature could reach -30 $\tccentigrade$ and the air flow rate could be adjusted steplessly.

Two chairs were placed 1 meter apart in the back center of the chamber for two participants in each experiment (Figure \ref{fig:fig1}.a). Cold air was ventilated into the chamber by fans installed at the upper part of the wall in front of the participants. A large piece of thick gray felt fabric was put up in front of the fan outlet to avoid direct draught. Six temperature and humidity recorders (Figure \ref{fig:fig1}.b) were deployed on both sides of the chairs at the height of 0.1 m, 0.6 m and 1.1 m, and two cardan anemometers (Figure \ref{fig:fig1}.c) were set at the height of 1.1 m. A black bulb thermometer (Figure \ref{fig:fig1}.d) was placed in the middle of the two chairs at the height of 0.6 m. (Parameters of the devices are shown in Table 1.) The temperature, relative humidity, black bulb temperature during the experiments were (mean ± std) -14.6 ± 0.4 $\tccentigrade$, 78.3 ± 4.8\%, -15.5$\tccentigrade$ ± 0.5 $\tccentigrade$ and air speed was controlled below 0.1 m/s at the occupant area.

\subsection*{Participants}

Ten healthy young university males were recruited as participants. They aged between 19 to 25 (22.0 ± 2.3) years old and their BMI was between 18.9 to 23.3 (21.5 ± 1.2) kg/m². None of them had previously participated in experiments conducted in extreme cold environments, and they were asked to avoid alcohol, caffeine, and drugs for at least 12 h before the experiment. Standard clothes were provided. Before experiments, participants put on three layers of clothes for both upper and lower body, in the order of long thick underwear, heating garment (developed by our team, will describe it in details below) and down coat and pants. Winter hats with masks, gloves, cotton socks and shoes were also provided (Figure 2.a). The thermal insulation of the ensemble was 1.67 clo measured by a thermal manikin (seat insulation included). All participants participated in a training session to learn the experiment procedure and regulations of the heating garments, and they consented to follow the experimental protocol, which required participants (1) not to engage in activities that required days of recovery (like blood donation, marathons, etc.) within two days before the experiment; (2) activities that could result in significant changes in metabolic rate (like sleep, eating, exercise, etc.) were not carried out one hour prior to the start of the experiment; (3) a regular daily schedule was well kept before the experiment.

\subsection*{Physiological measurement}

Skin temperatures of 12 body parts, i.e., forehead, chest, back, abdomen, waist, upper arms, forearms, hands, buttocks, thighs, calves and feet, were monitored by PyroButtons attached to skin test spots throughout the experiments (Figure \ref{fig:fig1}.e). A small piece of cotton insulation was attached to cover the bottom side of each sensor contacting with clothing in order to eliminate direct heat transfer from the heating surface of the heating garment. The ECG heart rate sensor was worn on the chest with high quality electrodes on its soft textile strap to ensure the device could record the heart rate with high accuracy (Figure \ref{fig:fig1}.f). Eight of ten participants voluntarily took the e-Celsius performance capsule to measure the gastrointestinal temperature, i.e., body core temperature (Figure \ref{fig:fig1}.g). Parameters of physiological measurement devices are shown in Table. \ref{table:1}.

\begin{table}[!ht]
    \centering
    \caption{Parameters of environmental and physiological measurement devices}
    \label{table:1}
    \begin{tabular}{lccc}
    \hline
        \textbf{device}& \textbf{type} & \textbf{precision} & \textbf{precisioncompany}\\ \hline  
        \textbf{Temperature and humidity recorder} & WSZY-1 & ± 0.5 $\tccentigrade$, ± 3\% (RH)  & \\ 
        \textbf{Cardan anemometer} & WWFWZY-1 & ± 0.05 m/s &TianJianhuayi \\
        \textbf{Black bulb thermometer} & HQZY-1 & ± 0.3 $\tccentigrade$ \\ 
        \textbf{Skin temperature sensor} & Pyrobutton-T & ± 0.1 $\tccentigrade$ &  Maxium\\ 
        \textbf{Heart rate sensor} & Polar H10 & - & Polar\\ 
        \textbf{Core body temperature capsule} & P022-Perf Pill & ± 0.1 $\tccentigrade$ & Bodycap\\ \hline
    \end{tabular}
\end{table}

\subsection{Intelligent personal comfort system based on interactive heating garment}

A novel intelligent interactive experiment system was developed, which consists of heating garments and user interface for participants and experiment management system for experimenters. It enables participants to adjust the surface temperatures of the heating garment and report their subjective feedback on tablets. Meanwhile it allows experimenters to monitor and control the experimental process instantly on the computer.

There are two pieces of heating garment, for upper and lower body parts. The heating units are modularized - each garment has 20 standard heating units, for each unit there is a PWM temperature control module with a temperature sensor with precision of 0.1 $\tccentigrade$ attached to the 20×20 cm² fabric with heating wires such that the heating surface temperature can be controlled to the set value within 20 seconds with precision of 1 $\tccentigrade$. The heating garment covers 9 body segments: chest, back, abdomen, waist, upper arms, forearms, buttocks, thighs and calves (4 pieces for thighs, 2 pieces for each other segments). Heating units for each segment could be controlled independently, thus any combinations in heating powers among the 9 body parts are allowed. 

\begin{figure}[h]
    \centering
    \includegraphics[width=16cm]{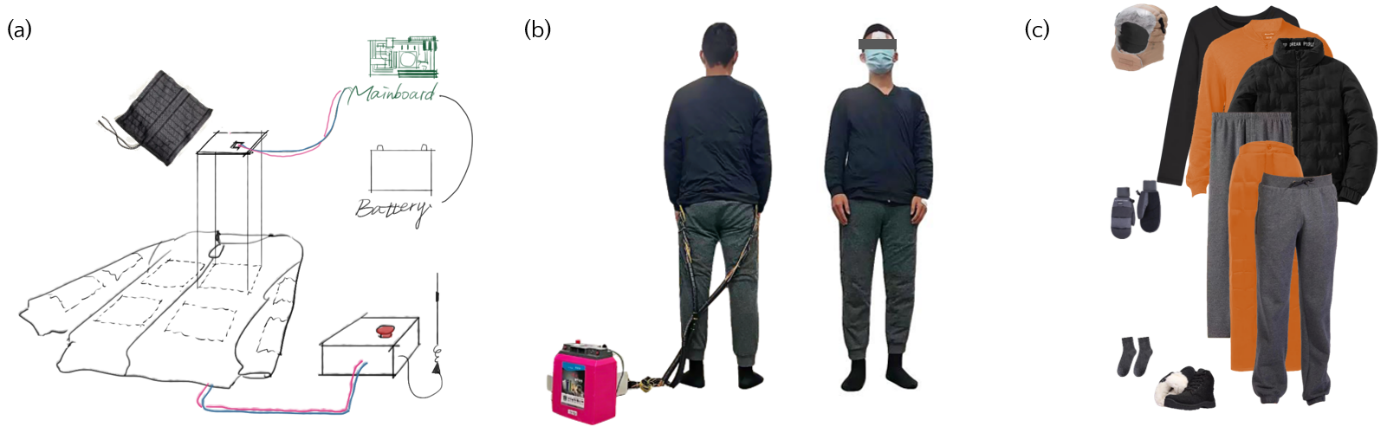}
    \caption{Heating garment used in the experiment. \small \textbf{a}. Basic structure of the heating garment (take the top as an example). \textbf{b}. Participant wearing the heating garment. \textbf{c}. Standard ensemble with the heating garment marked in orange. }
    \label{fig:fig2}
\end{figure}

Heating fabrics are fixed on the inner face of a soft, close-fitting sportswear (Figure \ref{fig:fig2}.a). Wires for power and communication are well-organized, which are fixed on the sportswear, collected together for upper and lower halves and finally connected to a control box with mainboard and communication antennas (Figure \ref{fig:fig2}.b). The weight of the garment is 1.15 kg for the wearing part and 1.82 kg with wires included. A red emergency button (Figure \ref{fig:fig2}.a) is set on the control box in the reach of participants to ensure the security of the experiment - the current will be switched off once the button is knocked down by the participants (when they feel uncomfortable). Parameters of the heating garment are shown in Table 2.


\begin{table}[]
\caption{Parameters of the heating garment}
\centering
\label{tab:table-2}
\begin{tabular}{lll}
\hline
\multirow{3}{*}{heating surface} & nominal parameter               & 5V / 2A       \\
                                 & size                            & 20 cm x 20 cm \\
                                 & maximum power                   & 35 W          \\ \hline
\multirow{4}{*}{heating garment} & number of heating units         & 20            \\
                                 & number of heating body segments & 9             \\
                                 & temperature control accuracy    & 1 $\tccentigrade$           \\
                                 & response time                   & < 40 s        \\ \hline
\multirow{2}{*}{safe limit}      & current of each heating cloth   & 3.2 A         \\
                                 & heating temperature             & 55 $\tccentigrade$          \\ \hline
mainboard                        & model specification             & STM32F103ZET6 \\ \hline
Temperature sensor               & model specification             & TMP275AIDGKR  \\ \hline
battery                          & Working voltage                 & 12 V          \\ \hline
\end{tabular}
\end{table}

\begin{figure}[h]
    \centering
    \includegraphics[width=16cm]{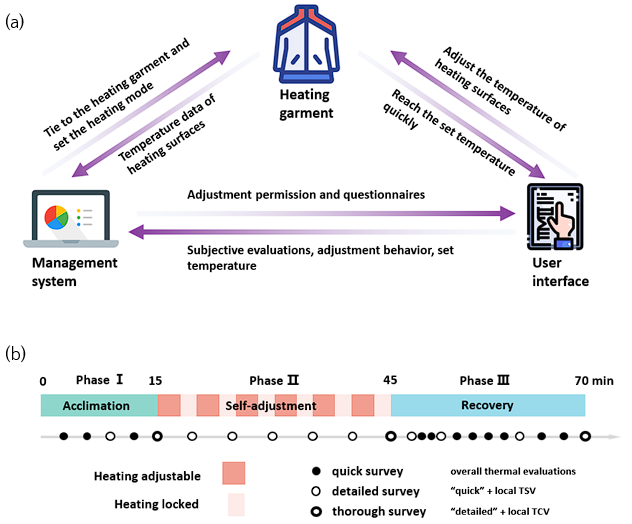}
    \caption{Constructure of the experiments. \small \textbf{a}. Logic of the experiment system.  \textbf{b}. Experiment procedure. }
    \label{fig:fig3}
\end{figure}

A web-based experimental platform was particularly designed for the experiment (Figure \ref{fig:fig3}.a). The user-interface performs three main functions for participants: (1) providing experiment information and instructions. The equipment id, the connection condition of the heating garment, the current experiment phase and timer are shown on the top. Detailed prompts are scheduled to provide information such as whether and how the heating garment could be adjusted at the moment. (2) enabling participants to adjust the surface temperatures of different body parts. When the heating garment is available for self-adjustment, participants can click the body segment they want to adjust and drag the slider to adjust the heating temperature. The color of the heated segments varies depending on the set temperatures and displays the color corresponding to the real-time temperature 20 seconds after adjustments. (3) collecting subjective feedback as scheduled. Questionnaires will pop up at the given time, and the data will be transferred to the experiment management system after participants click the submit button.

The experiment management system was designed to assist researchers in better organizing the experiment, including (1) standard for procedure definition: experiment information are given by a specific spreadsheet consisting of modularized experiment phases, in which the beginning and ending time are explicitly stipulated, and in each phase, the on-off state, adjustability and the heating temperature of the heating garment are set in order, while the content and popup time of each questionnaire is also arranged; (2) database for experiment results and instant monitoring module: data uploaded by the heating garment and participants can be monitored in real time and downloaded after the experiment.

\subsection{Experiment design}

The experiment is intended to investigate the distribution of thermal demands of different body parts and look for an effective way to improve the thermal comfort of people in cold environments.

Each participant participated in one round of experiment, which included three consecutive phases with a total duration of 70 minutes (Figure \ref{fig:fig3}.b): acclimation phase for 15 minutes, followed by self-adjustment phase for 30 minutes, followed by recovery phase for 25 minutes. Participants sat in the cryogenic chamber with local heating off during the acclimation phase in order to get rid of the influence of their thermal history outside. Then, they were enabled to adjust the surface temperatures of the heating garments on different body parts via their tablets during the self-adjustment phase. Lastly, local heating was set off again and participants sat in the chamber for another 25 minutes. The self-adjustment phase was the treated group, while the recovery phase was designed as a control group to evaluate how participants would respond without local heating. The acclimation phase was designed to be relatively short to simulate the ordinary process when participants transfer from indoors to outdoors.

The self-adjustment phase was divided into 6 short sections, each lasting 5 minutes: during the first 3 minutes of each section, participants could adjust the heating temperature of each body part freely via their tablets, while in the last 2 minutes, the adjustment options would be locked, i.e., the heating surfaces would maintain the heating temperature that participants set at their last adjustments. This mechanism was designed to ensure that when participants were asked to report their thermal evaluations at the end of each 5-minutes section, they were exposed to simultaneously recorded thermal conditions for at least one minute, considering that the heating up time in this dynamic process was not neglectable. In the case that after 6 times of adjustment, both surface temperatures and thermal perceptions approached stable, participants were considered to have found their most ideal heating temperature of each body part, namely the “stable temperature”. 

We collected overall and local thermal sensation (TSV, 9-point continuous scale), overall and local thermal comfort (TCV, 7-point continuous scale), overall thermal acceptance (TAcc, 6-point discrete scale) and self-reported thermal sensation change (dTSV, 5-point discrete scale). TSV, TCV and TAcc are standard metrics widely used in thermal comfort research, while dTSV is a new metric we introduce. A seemingly widely-ignored question is whether conventional thermal comfort survey scales (either 7-points or 9-points) are still reliable under extreme cold environments. Since such scales have a fixed lower bound, participants cannot voice their feeling of “getting even colder”. To tackle this challenge, we included an extra question to the questionnaire which asked “compared with your last voting, how does your feeling change at the moment?”. Participants could vote “feel colder” even if they have already voted “very cold” (-4) for their previous TSV survey.

In order to collect as much subjective data as possible, especially to capture the thermal response to the highly-dynamic environment, while without interfering with the experiment, three questionnaires in different lengths were designed (Figure \ref{fig:fig3}.b). Quick surveys (QS) consisted of overall TSV, TCV, TAcc and dTSV. Detailed surveys (DS) added local TSV of 13 body segments (face, neck, hands, feet, chest, back, abdomen, waist, upper arms, forearms, buttocks, thighs and calves) to QS, and thorough surveys (TS) furtehr added local TCV of the 13 body segments to DS. Participants were able to complete QS in 20 seconds or shorter, which allowed us to repeatedly collect their perceptions in very short intervals.

Before the experiment, the experimenter would first log into the experimenter end and match participants up with corresponding heating garments and upload experiment files. At the same time, participants would wear sensors used for physiological measurements and put on the clothing ensembles (including heating garments) (Figure \ref{fig:fig2}.c) outside the climate chamber. Then the experimenter would help them to connect the heating garments with the experiment system. After entering the climate chamber, participants could log in on the tablet through the experimenter’s guidance by an interphone and properly start the experiment.

\subsection{Ethic approval}

The experimental protocol was approved by the Institution Review Board of researchers’ institution (No. 20210062), which ensures the security and interests of the participants during the experiment. Before the start of the experiment, every participant signed the informed consent after being informed of the procedure and guidelines of the experiment and instructed how to deal with potential risks. Participants had the right to suspend or quit the experiment at any time if they felt uncomfortable physically or mentally. Experimenters kept paying close attention to the experiment process via experiment management system and monitoring video. Physiological testing equipment used in the experiment were all small, wireless, noiseless and non-damaging to the human body. The heating garment had multiple protection measures including emergency stop button, current monitoring and fuse. When participants adjusted the heating temperature over 40 $\tccentigrade$, they would be warned to avoid potential damage from raising the temperature too quickly. They were also instructed to stop the heating device whenever they feel undesired warm discomfort via the emergency button.\footnote{From our results, the maximum temperature set by the participants was 38 $\tccentigrade$, which falls within a safe range. Most participants find their stable heating temperature around 30 $\tccentigrade$. From the researchers' perspective, we set a relatively broad temperature control range for participants in order to maximize their flexibility and include as little assumption from the researchers as possible.}

\section{Result}
In this section, we report the results of our experiment. We first analyze the response of participants without local and personalized heating (acclimatization and recovery phases) as a baseline performance. Then we systematically introduce the findings in the self-adjustment phase, in the order of adjustment behaviors (garment surface temperatures), physiological reactions (skin temperatures) and subjective evaluations (questionnaire results), with an emphasis on the temporal, spatial and individual variances shown in above perspectives.

\subsection{Baseline case}

The heating garment was turned off during the acclimation phase and recovery phase. Thus, participants’ skin temperature kept declining (Figure \ref{fig:fig5}.a). The average skin temperature of 12 test points fell down 2.3 $\tccentigrade$ during the acclimation phase and 2.4 $\tccentigrade$ during the recovery phase. Due to direct contact with the chair, the buttock displayed the most obvious temperature reduction, which was 4.0 $\tccentigrade$ and 5.0 $\tccentigrade$ separately for the two phases. The extreme cold environment also brought strong impact to subjective evaluations. At the end of the acclimation period, TSV, TCV and TAcc dropped to -1.7, -1.4 and -1.1 separately. In the recovery phase, subjective evaluations had a sharp decrease, which returned to the preheating level (the end of acclimation) in just 4 minutes, and continued to decline till the end of the experiment. Eventually, the TSV, TCV and TAcc reached -3.2, -2.4 and -2.5, indicating a thermal status even worse than “cold”, “uncomfortable” and “unacceptable”. It is also noteworthy that although overall TSV declines slower in the last 15 minutes, throughout the recovery phase, most participants keep reporting ``colder" in self-reported sensation change (dTSV). This suggests that including self-reported sensation change is necessary to understand the dynamics under extreme environments.

\subsection{Adjustment behaviors}

\begin{figure}[h]
    \centering
    \includegraphics[width=16cm]{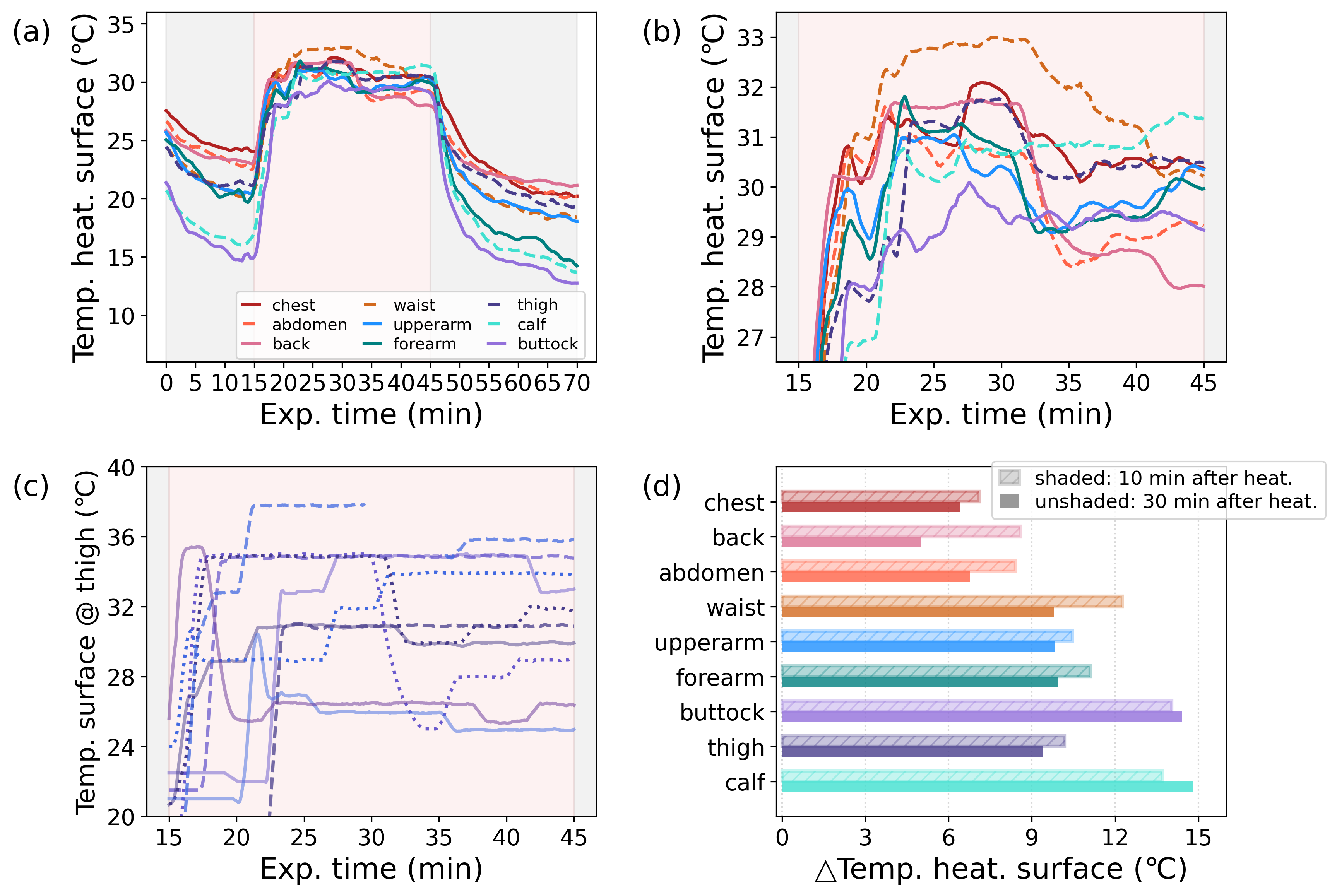}
    \caption{Temperatures of heating surfaces. \small \textbf{a}. Temperature of heating surfaces at each body segment throughout the experiment period. \textbf{b}. A zoom-in view of surface temperature focusing on the heating period.  \textbf{c}. Temperature of heating surfaces at thigh, controlled by different participants during the heating period. \textbf{d}. Temperature changes of heating surfaces at each body segment when heating (45 min), compared with stable temperature without heating (15 min). Bars with “//” hatches for 10 minutes for heating (25th minute); fully filled bars for 30 minutes after heating (45th minute). \textbf{Note}. Measured temperatures, instead of assigned temperatures are used. For body segments covered by more than one heating cloth, the average temperatures are used. In \textbf{a, b, c,} light-red shaded background marks the experiment period when local heating was on and personalized adjustment was allowed. In \textbf{a, b, d,} each line represents the average temperature of all participants at the corresponding body segment; in c, each line represents one particular participant.}
    \label{fig:fig4}
\end{figure}

All participants actively adjusted the surface temperatures of different body parts during the self-adjustment phase. Surface temperatures were monitored from the temperature sensors in each of the heating unit. Figure \ref{fig:fig4} (a-d) illustrates the temporal, spatial and individual patterns of participants’ adjustment behaviors reflected on the changes of heating garment surface temperatures. Once participants were permitted to turn on local heating and personally set the target surface temperatures, they immediately set the heating temperatures to a relatively stable condition within the range of 29 - 33 $\tccentigrade$, lifting the surface temperatures averagely by 10.1 $\tccentigrade$ compared with that of the end of acclimation period. They also kept fine adjustments throughout the self-adjustment phase (Figure \ref{fig:fig4}.a).  

Meanwhile, surface heating temperatures varied among different body segments. The difference between calves and back (body segment with the highest and lowest average surface temperature by the end of heating phase respectively) were up to 4 $\tccentigrade$, which indicates considerable thermal demand variances among body segments (Figure \ref{fig:fig4}.b).  Compared with the beginning of the heating phase, surface temperature increases in calves, buttocks and waist were most significant, while changes in chest, abdomen and back were relatively more moderate. It is also notable that after 10 minutes heating, participants slightly lowered the surface temperatures of most segments until the end of the heating phase, except calves and buttocks, and the retracement on waist and back reached around 3 $\tccentigrade$ (Figure \ref{fig:fig4}.d). At the beginning of the heating phase, participants had urgent heating needs both physiologically and psychologically and tended to “overheat” compared to their stable needs, and also enjoyed thermal pleasure caused by sensation changes. The cold discomfort gradually eased during the heating phase and local thermal demands approached a balance driven by thermal equilibrium, that is, for most body parts, especially for torso parts with high heat generation and good thermal insulation, heating surface temperatures could be lowered, while for calves (more exposed to the ambient environment) and buttocks (high heat conduction coefficient through the plastic mesh chair), heat loss was still the dominant part. In general, limbs were of higher thermal demands than torso parts.

Inter-individual differences also played an unneglectable role in a holistic understanding of variances in thermal demands. Exposed to the same environment, even for the same segment at the same time, heating surface temperatures set by different participants differed a lot from each other. Taking thighs as an example (Figure \ref{fig:fig4}.c), its stable temperature difference of two participants could reach up to 12 $\tccentigrade$, almost 3 times of the difference in body segments of an “average participant”, which suggests that it may be risky to assume a universal heating distribution pattern that everyone would accept, and the flexibility of personalized adjustment is what we need instead.

\subsection{Physiological reactions}

\begin{figure}[h]
    \centering
    \includegraphics[width=16cm]{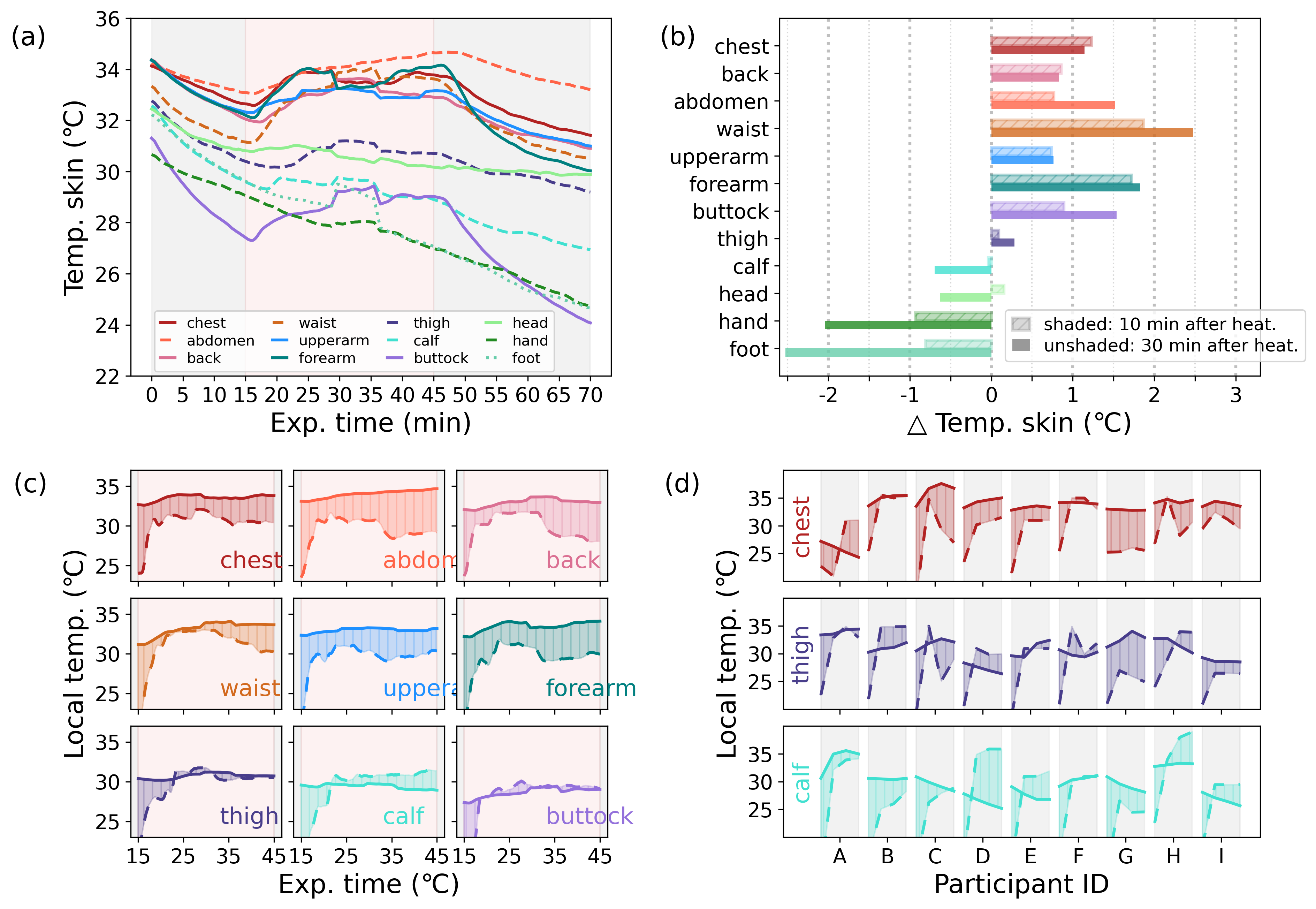}
    \caption{Skin temperatures. \small \textbf{a}. Skin temperature at each body segment throughout the experiment period. \textbf{b}. Skin temperature changes at each body segment when heating, compared with the temperature at the 15th min. Bars with “//” hatches for 10 minutes for heating (25th minute); fully filled bars for 30 minutes after heating (45th minute).  \textbf{c}. Local skin temperatures (solid lines) and temperature of corresponding heating surfaces (dashed lines) during the heating period. \textbf{d}. Local skin temperatures (solid lines) and temperature of heating surfaces (dashed lines) at chest, thigh and calf of each participant during the heating period. \textbf{Note}. In \textbf{a, b, c,} each line/bar represents the average temperature of all participants at the corresponding body segment; in \textbf{d,} each line represents one particular participant.}
    \label{fig:fig5}
\end{figure}

Skin temperatures directly indicate thermal status and influence thermal sensations, thus acting as a significant physiological criterion for evaluating the performance of local heating.  With surface heating temperatures adjusted, increases in skin temperatures of the heated segments followed instantly at the beginning of the self-adjustment phase, and most of them witnessed a substantial increase throughout the phase, except for calves (Figure \ref{fig:fig5}.a). Compared to the end of the acclimation period, skin temperatures of torso parts (including buttocks) averagely rose 0.8 - 2.5 $\tccentigrade$ by the end of the heating period, among which waist was the body segment with the largest skin temperature lift as well as 
the highest stable temperature (Figure \ref{fig:fig5}.b). For limbs directly heated, upper arms and forearms’ skin temperatures increased, while calves and thigh’s decreased, since trousers were less close-fitting thus cold air infiltrated into the interlayer between heating garment and underwear. For those unheated body parts, i.e., head, hands and feet, their skin temperatures kept dropping - decreased 0.7 $\tccentigrade$ to 2.7 $\tccentigrade$ within the 30-minute self-adjustment phase.

When comparing skin temperatures with surface heating temperatures, we found that for the eight body segments that achieved skin temperature rise through the heating period, seven of them showed surface temperature retracement during their adjustments (Figure \ref{fig:fig5}.c). Calves, as the only heated part whose skin temperature kept declining, were treated with increasing heating surface temperatures throughout the adjustment phase. By comparison, we also found that in general, the stable heating surface temperature would not exceed regular skin temperatures, which is 30 - 34 $\tccentigrade$.

Again, it was not surprising to find that although some patterns were seemingly suggested from those “averaged” results, when looking at the individual level, these patterns usually could not apply. Participants showed various reaction patterns with diverse setting temperature and adjustment timing for the same body segment - however, an abstract rule held: participants adjusted adaptively attempting to mantain their skin temperatures at the stable temperature.

The mean heart rate during the self-adjustment phase is 73 bpm, which is lower than 76 bpm during the acclimation phase but slightly higher than 72 bpm during the recovery phase. The heating showed an effect of reducing participants’ metabolic and this effect could last in a short period. There were eight participants voluntarily participated in the test of core temperature. The mean core temperature was basically stable during the experiment, which was 37.5 $\tccentigrade$ for the start of the experiment and 37.3 $\tccentigrade$ for the end (the normal range of individuals’ core body temperature fluctuates within 1 $\tccentigrade$). Local heating did not significantly increase or decrease participants' core temperature. 

\subsection{Subjective evaluations}

\begin{figure}[h]
    \centering
    \includegraphics[width=16cm]{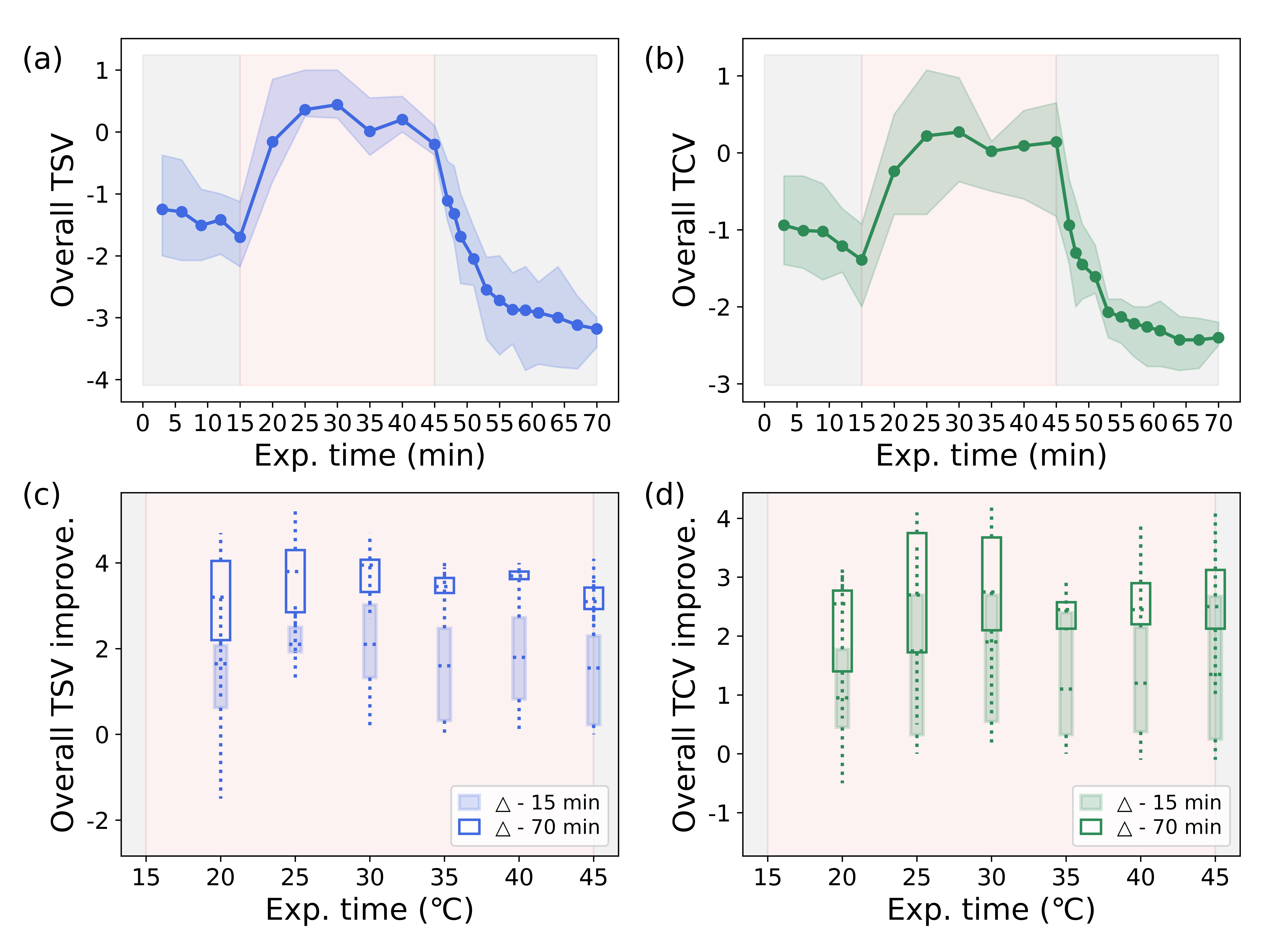}
    \caption{Overall thermal sensation (TSV) and comfort (TCV) votes. \small \textbf{a., b.} Overall TSV / TCV throughout the experiment period. Each dot marks one subjective survey. The line represents the average vote of all participants and the shadow area shows the range of 25th and 75th percentiles. \textbf{c., d.} Overall TSV / TCV improvement: the last vote during the heating period (45th minute) compared to that at the end of acclimation (15th minute, hollow box) and the end of the experiment (70th minute, shaded box) during the heating period.}
    \label{fig:fig6}
\end{figure}

Participants’ active adjustments effectively improved their subjective thermal perceptions (Figure \ref{fig:fig6}). The mean overall TSV at the end of the self-adjustment phase was raised to -0.2, i.e., the microclimate was regarded as thermal neutrality by most participants, which was 1.5 unit higher than that at the end of acclimation ($p<0.05$), realizing the upgrade from “cold” to a thermal neutral stage and the TSV during the self-adjustment phase was kept around 0.1. 

Accompanied with the feeling of warmth, adaptive local heating also made participants feel more comfortable. TCV curves throughout the self-adjustment phase showed a parallel trend with TSV curves, and on average, overall TCV was lifted from -1.4 at the start of self-adjustment to 0.1 by the end of the phase ($p<0.05$). 

\begin{figure}[h]
    \centering
    \includegraphics[width=16cm]{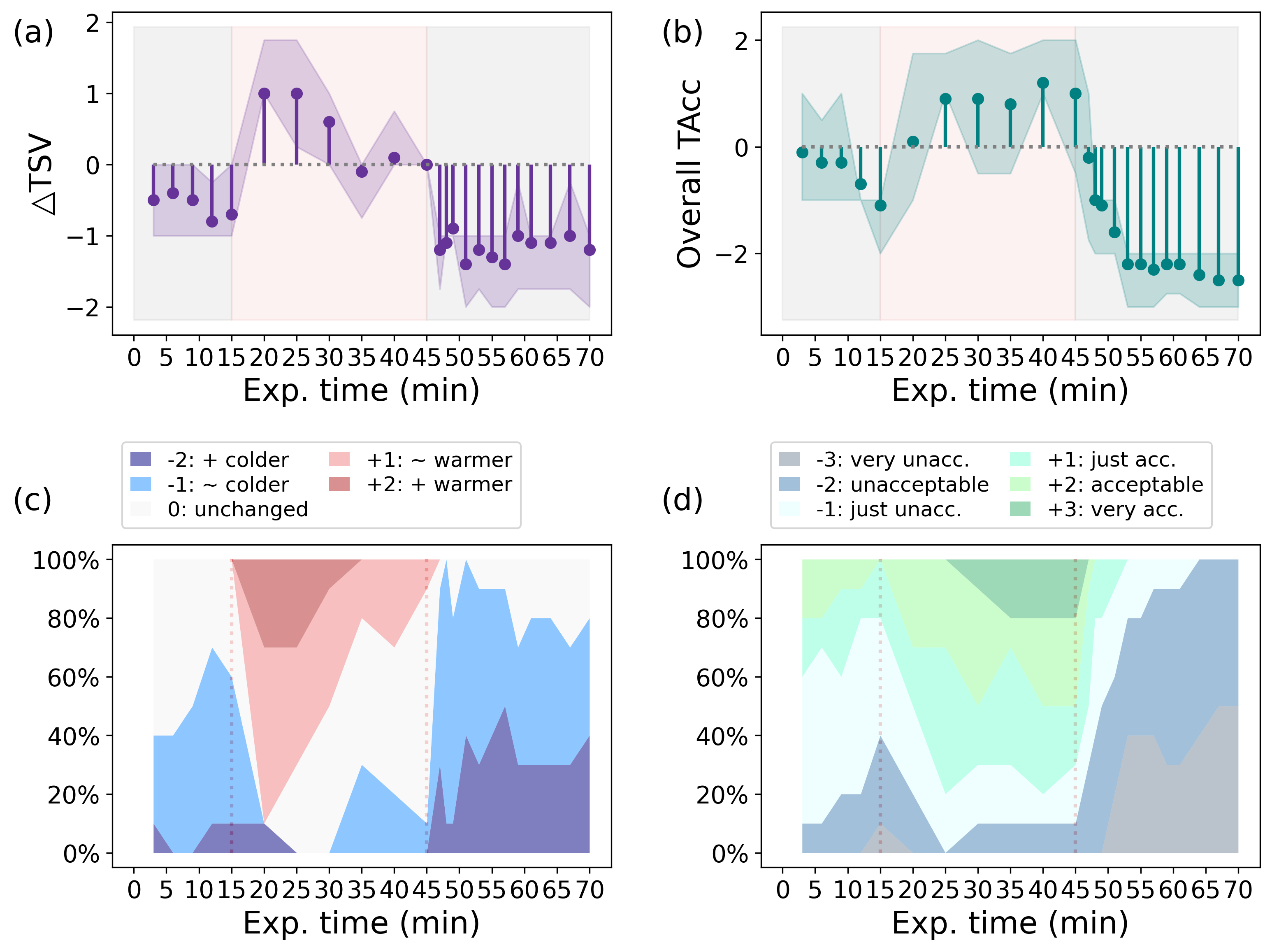}
    \caption{Thermal acceptance and thermal preference. \small \textbf{a}. Self-reported sensation change (dTSV) throughout the experiment period. \textbf{b}. Overall TAcc throughout the experiment period. \textbf{c}. dTSV throughout the experiment shown in percentage. \textbf{d}. Overall TAcc throughout the experiment period shown in percentage. \textbf{Note}. In \textbf{a} and \textbf{b}, the dot-line represents the average vote of all participants and the shadow area shows the distribution between the upper and lower quartiles.}
    \label{fig:fig7}
\end{figure}

Despite heterogeneous thermal preferences of different participants, after 15 minutes of adjustment, most of them reached stable thermal neutrality, reporting “no change” in self-reported sensation change for the remaining 15 minutes (Figure \ref{fig:fig7}.a, c). Once the heating garment was turned off, participants obviously felt a thermal sensation decrease and became colder and colder till the end of the experiment. TAcc rose rapidly when participants could adjust temperatures of heating surfaces (Figure \ref{fig:fig7}.b). During the self-adjustment phase, overall acceptance (TAcc $\ge$ +1) could reach 80\% (Figure \ref{fig:fig7}.d), indicating that the heating garment could enable 80\% of participants to accept the -15 $\tccentigrade$ environment. 

\begin{figure}[h]
    \centering
    \includegraphics[width=16cm]{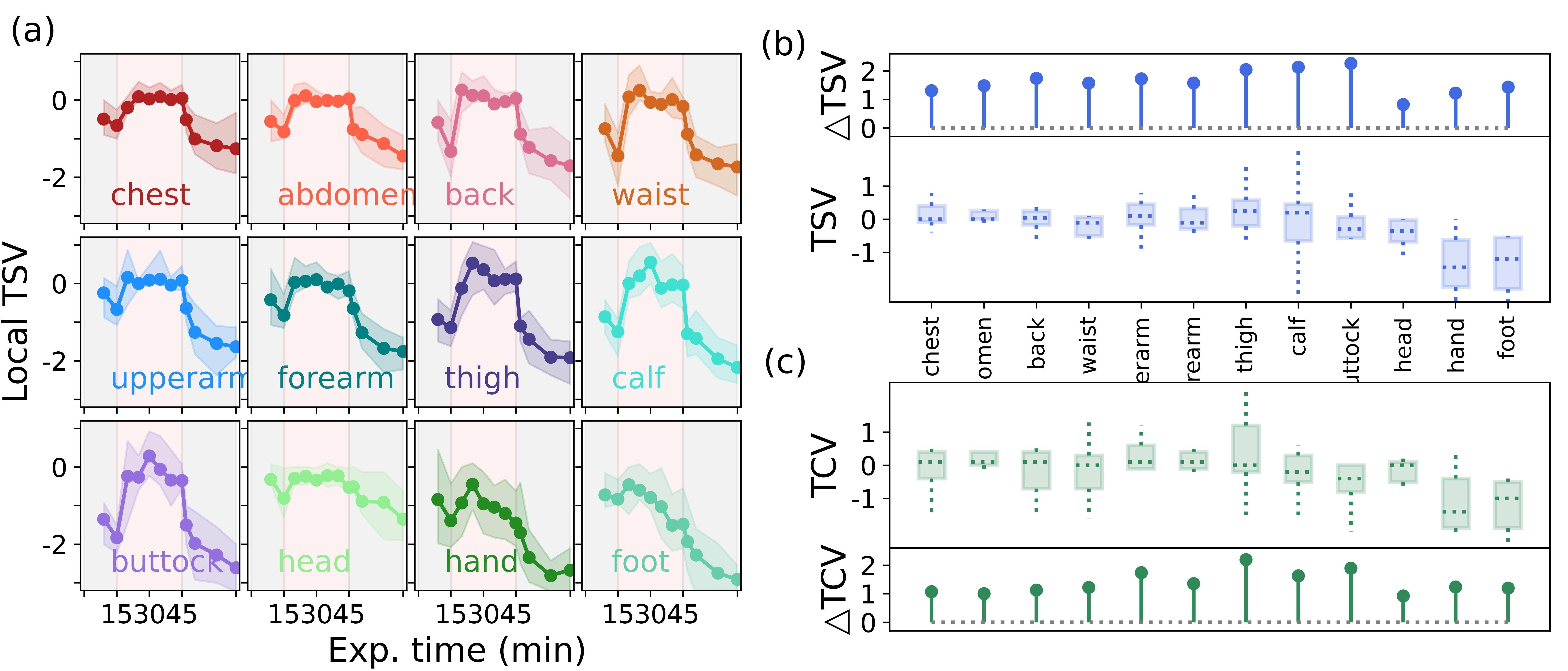}
    \caption{Local TSV and local TCV. \small \textbf{a}. SLocal TSV of 12 body segments throughout the experiment period. \textbf{b., c.} stem plot: Local TSV / TCV improvement (the vote at the end of the self-adjustment phase at 45 min compared to the vote at the end of the experiment at 70 min) and boxplot: local TSV / TCV distributions at the last vote of the self-adjustment phase (45th minute).}
    \label{fig:fig8}
\end{figure}

Local thermal comfort was also considerably improved during the self-adjustment phase. Self-adjustment could satisfy the thermal demands of every body segment being covered by the heating garment (Figure \ref{fig:fig8}.a). Local sensation from all those heated body parts reached to a sensation near neutral during the self-adjustment phase, improved effectively with 1 -- 2 units compared to the end of the experiment (Figure \ref{fig:fig8}.b). TSV of body parts without heating also had a slight increase at first but quickly went down as the case of hands and feet, while the head part became an exception that its TSV was successfully stabilized during the self-adjustment phase. With self-adjustment, inter-individual differences in overall and local thermal perceptions were narrowed (Figure \ref{fig:fig8}.a), which is another convincing evidence that participants themselves were able to respond appropriately to their diverse thermal demands. 

\section{Discussion}
\subsection{Comparison with existing research}

We compare our results with some existing research on local thermal comfort, mainly \cite{deng_effects_2019}, \cite{zhang_thermal_2010}, \cite{luo_high-density_2020} and \cite{zhang_review_2015}. 

Deng et al \cite{deng_effects_2019} is among the very limited research conducted in similar environments, in which participants were provided with heating cushions, footpads and heating gloves for local heating. In their results, even the heated body segments failed to maintain thermal neutral locally, and the TSV kept declining during heating. At the 55th minute, the TSV for hands, feet and buttocks reached -2.00, -1.63 and -1.00. In our study, TSV of all body segments under heating reached stable during the self-adjustment phase and TSV of the head, which was not covered by heating surfaces, was also stabilized. This may illustrate that heating only three areas like hands, feet and buttocks are not enough to improve overall thermal comfort to neutral level in extreme cold environments. The probable reason might be that the combination of heating body segments is not optimal and the self-control is limited to only three given parts.

Zhang et al \cite{zhang_thermal_2010} established a comprehensive model of local thermal comfort based on data from laboratory experiments with cooling/heating air sleeves as stimuli devices. Similar to the finding in Deng’s study \cite{deng_effects_2020}, there exists discrepancy between our results and prediction via Zhang’s model. Our experiment data may further help correct / generalize Zhang’s model to extreme cold environment, where few data were collected when the model was originally developed. 

Luo et al \cite{luo_high-density_2020} tested body thermal sensitivity at a very fine scale, however, as Ju et al \cite{ju_development_2020} questioned, it is still not very clear how to interpret the resourceful physiological data into guide for PCS design (e.g., whether participants will prioritize heating body parts of low sensitivity or high sensitivity). Our results can provide some insights on this question. For example, the calves have the highest thermal demand (14.9 $\tccentigrade$) and they were evaluated to be the part which has the lowest thermal sensitivity to heating stimuli.

Further, Zhang et al \cite{zhang_review_2015} propose an index “corrective power” (CP) to evaluate the performance of different types of PCS. By its original definition, CP is defined as “difference between two ambient temperatures at which the same thermal sensation is achieved - one with no PCS (the reference condition), and one with PCS in use”. In our experiment, the ambient temperature was -15 $\tccentigrade$, and with personalized heating, overall TSV was maintained around 0, the same as if participants were in a uniform environment of 20 $\tccentigrade$, so CP of our devices was 35 $\tccentigrade$. It doesn't necessarily mean our system is, for instance, 3 times more effective than a device that works in 10 $\tccentigrade$ and has a CP of 10 $\tccentigrade$. However, our system is among very few personalized devices which can correct an extreme cold environment into an almost acceptable one.

Lastly, it’s also noteworthy that in some previous research such as \cite{yang_study_2020} and \cite{zhang_effect_2007}, it is reported that although TSV is improved with higher power of PCS, overall TCV doesn’t improve or even worsen due to large differences in local thermal sensations. In our experiment, as shown in Figure.~\ref{fig:fig6}, the overall TCV curve shares the same trend with TSV improvement. It is because in our experiment, participants can adjust heating surfaces covering a large proportion of their body instead of only one or two body parts, and they tend to set surface temperatures uniform thus avoiding large local sensation differences. It can be concluded that local heating with our system is more effective in terms of improving participants’ comfort.

\subsection{Adaptive adjustments}

As shown in the results part, thermal demands in cold environments are highly heterogeneous and dynamic, and the variances can be summarized from temporal, spatial and individual levels. When looking at the mean values, it seemingly suggests some explicit rules, for instance, torso parts are less in need of local heating than limbs. However, only recognizing these rules are insufficient to meet everyone’s thermal demands, because variances are almost everywhere (e.g., see Figure.~\ref{fig:fig4}.c, Figure.~\ref{fig:fig5}.d).

We propose a thermo-physiological mechanism to summarize all phenomena seen from our experiments abstractly as “adaptive adjustments”. Adaptive adjustments basically mean participants would make adjustments on local comfort devices based on their thermal perceptions and expectations. For instance, one would raise the local heating power up on some body part if she feels cold on that segment / the segment caused stronger discomfort. Different individuals may have different sensitive body segments, so their adjustments may not agree with each other on the same segment at the same time. However, if we look at the relationship among adjustment behavior, physiological reaction and thermal perception, certain agreements are shown. For instance, in Figure 5.c, although different body segments have different thermal-physiological characteristics and thermal demands, the general pattern is for those whose skin temperatures increase and approach stable temperatures, their heating surface temperatures showed retracement, while if skin temperature keep decreasing, participants would set surface temperatures even higher with the hope to compensate the heat loss. Another example can be found in Figure \ref{fig:fig5}.d, where both skin temperatures and adjustment behaviors of participants are quite diverse, but their relationship is rather clear and consistent.

Adaptive adjustment is a conceptual framework on understanding participants' interaction with the environment and their dynamics in thermal status. Though it does not directly quantify the performance of different PCS devices, a general principle of PCS design, especially under extreme environments, would be providing more flexible adjustment options. Most PCSs only include adjustment options for total power \cite{pasut_energy-efficient_2015}, \cite{arghand_individually_2022}, however, our results suggest that it’s vital to take variances in thermal demand distributions into account.

\subsection{Limitations}

One of the main limitations of our work is the representativeness of our participants. The researchers recognize that the sample size is below a typical size of thermal comfort research due to the difficulty in conducting experiments in extreme cold environments with complicated experimental devices and procedures. For scientific soundness, we consider the following justifications: First, our experiment was conducted in the well-controlled artificial climate chamber. Further, participants were from single sex and similar age group. Thus, many exogenous influential factors were eliminated. Second, the effect size of local heating in extreme cold environments was supposed to be large. Further, since our experiment was designed to be \emph{within-subjects}, as long as the effect of intervention is consistent, individual difference itself does not matter much. Thirdly, as a golden rule, we examine the statistical significance of our results. Our main conclusions in thermal comfort improvement are statistically significant ($p<0.05$).

There is evidence from previous research that sex, age, and adaptation factors may make a difference in the results. Females are more likely to feel colder than males in the same cooling protocol \cite{kaikaew_sex_2018}, elderly’s vascular regulation ability usually degrades thus their thermoregulatory responses are delayed compared to the young \cite{maeda_seasonal_2005}, and subjects who lived in cold indoors had a milder physiological reaction than those lived in neutral-to-warm wintertime indoor climates \cite{luo_indoor_2016}, \cite{cao_too_2016}. Metabolic rate also plays a role in thermal comfort. However, since our heating garment can satisfy the requirements under sedentary conditions (the local heating need is more urgent), it should be expected to be applicable for exercise scenarios. A follow up research investigates the performance with higher metabolic rates, and concludes that local heating can still efficiently improve participants thermal comfort \cite{LI2023109798}.
We would highly appreciate follow-up research to study the thermal demands of some specific group of people. However, we would like to argue that the core of our research is not to find those “static” conclusions, on the contrary, we highly emphasize the heterogeneous nature of thermal demands and suggest more flexible control options to accommodate different users. In such a sense, we are happy to see any “contradictory” findings with different groups of participants because these findings actually endorse our argument aforementioned.

In the experiment, the battery we used to supply power to the heating garment was not portable. In the design of the heating garment, we prioritized the flexibility, accuracy and response time of our heating garment control, and provided redundant power for users so that they can explore a wide range of heating combinations and adjust to exactly what they prefer. For real application, batteries with smaller capacity and nominal power should be sufficient. Since we mainly increase the complexity in control logistics, but no substantial increase in energy consumption, and there are some off-the-shelf products showing that personalized heating can be supported by portable batteries, portability should not be a major concern for potential commercialization of the heating garment.

Another limitation is although we design the experiment with the ambition to make it a research paradigm for local comfort research, our conclusions in this paper only hold for local heating in cold environments. When it comes to the local cooling side, we look forward to similar research designs which enhance the flexibility of local cooling combinations and personal controls, but there is certainly a lot of specific system design work required. One of the hints might be that we need to carefully consider the appropriate heat transfer approach - conductive is effective for local heating, but convective may be more suitable for local cooling.

\subsection{Future work}

Based on our current work and current system, we are considering some extensions for future research. One important question is how to allocate the heating power when there's a constraint on the total power for each participant. Such a constraint is related to many aspects of product design, for instance, the weight of batteries. In our presented work, participants have unlimited total heating power, and they tend to finally allocate the power uniformly on different body parts. However, when total power is limited, especially when it is not possible to set all surface temperatures as desired, would participants lower the surface temperatures uniformly, or would they overheat some specific body parts while sacrificing others? We look forward to future work on this question.

Another question is whether we can further make use of the dynamics of thermal demands. To be more specific, inspired by the phenomenon known as persistence of vision, we would like to explore dynamic heating by alternating local heating on different body parts, which would be much more energy-efficient. Our hypothesis is that due to the thermal mass of the garment as well as perceptional inertia, participants would feel warm shortly after local heating turns off. In Figure \ref{fig:fig6}.a, the overall sensation had a significant drop from neutral to -2 in 5 min when all the local heating units were turned off.  However, the effect might be different if at any moment, only part of the heating surfaces are off while others are on, and each surface is turned on for part of time during, say every three-minute slot.

\section{Conclusion}
The presented research reports systematic evidence on personalized adaptation’s effect on neutralizing individual, spatial and temporal thermo-physiological variances from experiments of novel design. Specifically, main conclusions are listed as follows:

\begin{itemize}
    \item Active heating could significantly and stably improve participants’ thermal satisfaction in extreme cold condition (-15 $\tccentigrade$). Participants TSV were maintained as “neutral” (-0.2) in an otherwise “very-cold” (-3.2) environment;
    \item Thermo-physiological variances in individual, spatial and temporal and their cross-effects were evaluated, revealed from sin temperature measurements and subjective responses. Spatially, limbs were more in need of heating than torso parts, whose difference in stable heating surface temperatures can be up to 4 $\tccentigrade$. Temporally, behaviors like overheating and retracement and psychological response like sensation overshoot were observed. Individually, even for the same segment, interpersonal differences could be up to 10 $\tccentigrade$.
    \item Personalized adaptation on local heating powers effectively neutralized above variances, i.e., almost every participant enjoyed a similar level of comfort although they had diverse requirements. More than 80\% of participants perceived their final thermal status “acceptable”. Individual variances in local TSVs during the self-adjustment phase were also significantly narrowed.
    \item “Adaptive adjustment” is proposed as a general rule to comprehensively understand aforementioned variances and participants’ responses. It’s suggested that practitioners should provide more flexible adjustment options in PCS design.
\end{itemize}

\section*{Data availability}

We will open source all the raw experiment data upon being accepted.

\section*{Declaration of interests}

The authors declare that they have no known competing financial interests or personal relationships that could have appeared to influence the work reported in this paper.

\section*{Acknowledgments}

This work was sponsored by Beijing Nova Program (No. Z191100001119051) and National Natural Science Foundation of China (No. 52078270 and No. 51838007). The authors would like to express their sincere gratitude to help from members in Key Laboratory of Refrigeration Technology, Tianjin University of Commerce and all the participants involved in the experiment. Some preliminary results of this paper have been presented on the 17th International Conference of the International Society of Indoor Air Quality \& Climate (Indoor Air 2022, Kuopio, Finland).

\section*{Credit author statement}
\textbf{Yi Ju}: conceptualization, methodology, formal analysis, investigation, software, writing - original, writing – review \& editing; \textbf{Xinyuan Ju}: methodology, formal analysis, investigation, software, writing - original; \textbf{Hui Zhang}: conceptualization, supervision, writing – review \& editing; \textbf{Bin Cao}: conceptualization, supervision, funding acquisition, writing – review \& editing; \textbf{Bin Liu}: supervision, writing – review \& editing; \textbf{Yingxin Zhu}: supervision, writing – review \& editing

\appendix
\newpage
\section{User interface demo}
\begin{figure}[h]
    \centering
    \includegraphics[width=12cm]{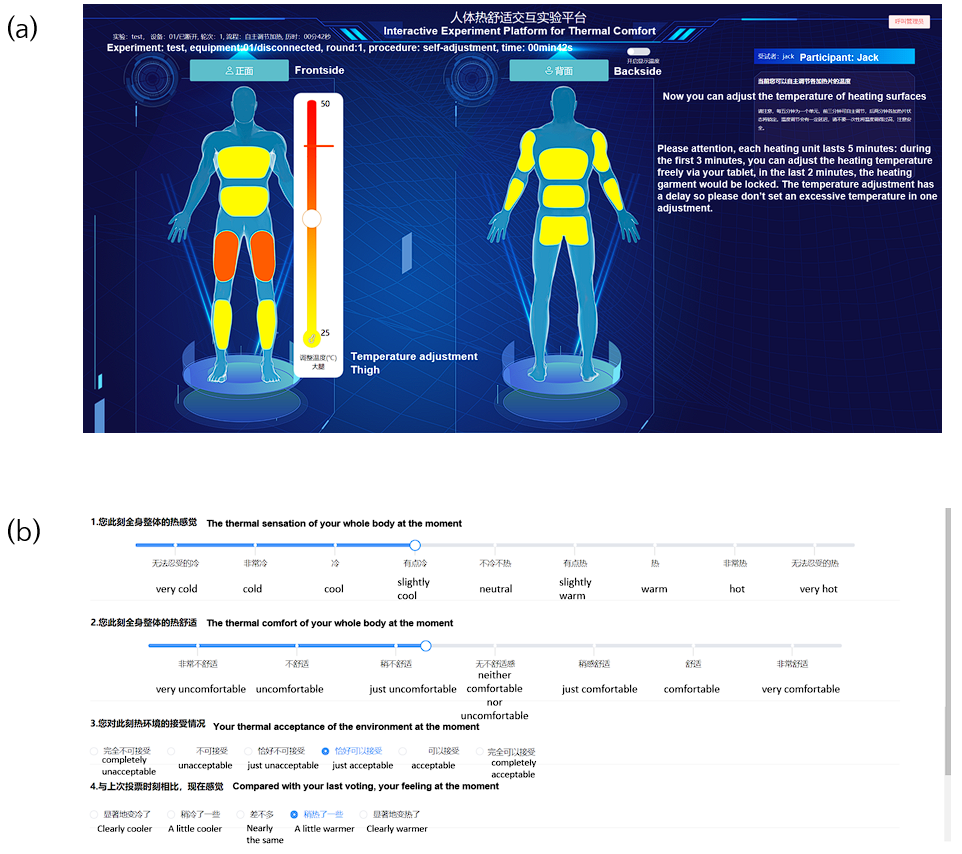}
    \caption{Screenshot of user interface. \small \textbf{a}. interface where users can adjust garment surface temperatures of each particular body part. \textbf{b}. users fill questionnaires in the interface}
    \label{fig:fig9}
\end{figure}

\section{Questionnaire used in the experiment}
\textbf{Q1-1} The thermal sensation of your whole body at the moment

\includegraphics[width=12cm]{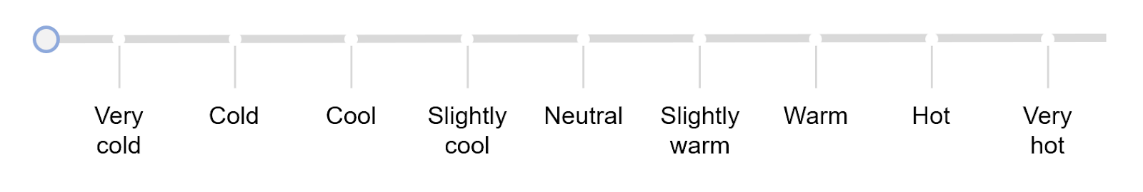}

\textbf{Q1-2} The thermal comfort of your whole body at the moment

\includegraphics[width=12cm]{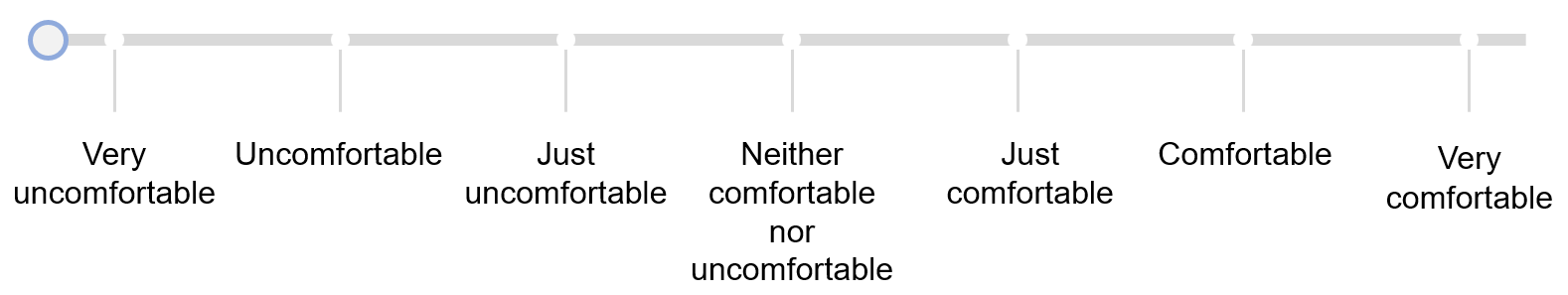}

\textbf{Q1-3} Your thermal acceptance of the environment at the moment

\begin{tabular}{lll}
$\square$~ Completely unacceptable 
& $\square$~ Unacceptable 
& $\square$~ Just unacceptable\\
$\square$~ Just acceptable 
& $\square$~ Acceptable 
& $\square$~ Completely acceptable
\end{tabular}

\textbf{Q1-4} Compared with your last voting, your feeling at the moment

\begin{tabular}{lll}
$\square$~ Clearly cooler 
& $\square$~ A little cooler 
& $\square$~ Nearly the same\\
$\square$~ A little warmer 
& $\square$~ Clearly warmer 
\end{tabular}

\textbf{Q1-5} Assuming you are watching the Winter Olympics live, the one that best matches your current feelings is

\begin{tabular}{ll}
$\square$~ I can’t wait to leave 
& $\square$~ I can only stay here up to 20 minutes\\ 
$\square$~ I can barely stay here for 40 minutes
& $\square$~ It’s not too difficult for me to stay here for 60 minutes\\ 
\multicolumn{2}{l}{$\square$~ I would keep watching as long as the game is fascinating since I am satisfied with the thermal condition }
\end{tabular}

\textbf{Q2-1*} The thermal sensation of your \texttt{<body part>} at the moment

\includegraphics[width=12cm]{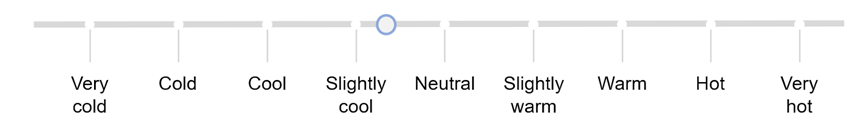}

\textbf{Q2-2*} The thermal comfort of your \texttt{<body part>} at the moment

\includegraphics[width=12cm]{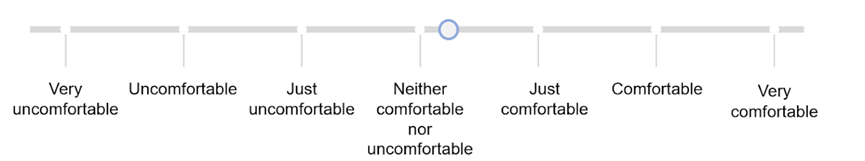}

\textbf{Q2-3} Please select the part of your body that is currently experiencing strong discomfort (select up to 4 and start with the most discomfortable one)

\begin{tabular}{lllllll}
$\square$~Face & $\square$~Neck &  $\square$~Chest &  $\square$~Abdomen	&  $\square$~Back &  $\square$~Waist	\\
$\square$~Upper arm	& $\square$~Forearm & $\square$~Hand	 & $\square$~Buttocks &  $\square$~Thigh & $\square$~Calf	& $\square$~Foot
\end{tabular}

\textbf{*} 13 body segments are surveyed in Q2-1 and Q2-2: chest, back, abdomen, waist, upper arms, forearms, buttocks, thighs, calves, feet, hands, neck and face.

\section{Result summary}

\begin{landscape}
\begin{table}[]
\caption{Thermal response of the end of heating / the baseline case (\textit{**: $p<0.05$, *: $p< 0.1$})}
\label{tab:appex-C}
\begin{tabular}{lcccc|cccc}
\hline
\multirow{2}{*}{\textbf{segments}} &
  \multicolumn{4}{c|}{\textbf{45th min (the end of self-adjustment phase)}} &
  \multicolumn{4}{c}{\textbf{70th min (the end of recovery phase)}} \\ \cline{2-9} 
 &
  \textbf{surface temp} &
  \textbf{skin temp} &
  \textbf{TSV} &
  \textbf{TCV} &
  \textbf{surface temp} &
  \textbf{skin temp} &
  \textbf{TSV} &
  \textbf{TCV} \\ \hline
\textbf{overall}   & -          & -          & -0.2 ± 0.8** & 0.1 ± 1.3**  & -          & -          & -3.2 ± 0.5 & -2.4 ± 0.3 \\
\textbf{chest}     & 30.5 ± 2.6 & 33.8 ± 3.5 & 0.1 ± 1.0**  & 0.1 ± 0.8**  & 20.0 ± 2.7 & 31.4 ± 3.7 & -1.3 ± 1.0 & -1.0 ± 0.7 \\
\textbf{back}      & 28.7 ± 3.4 & 32.9 ± 2.6 & 0.0 ± 1.0**  & -0.2 ± 1.2*  & 20.6 ± 2.1 & 30.9 ± 1.7 & -1.7 ± 0.8 & -1.3 ± 0.7 \\
\textbf{abdomen}   & 29.5 ± 2.6 & 34.6 ± 1.6 & 0.0 ± 0.8**  & 0.0 ± 1.2*   & 19.5 ± 3.7 & 33.2 ± 1.5 & -1.5 ± 0.8 & -1.0 ± 0.8 \\
\textbf{waist}     & 30.7 ± 3.8 & 33.6 ± 2.5 & -0.2 ± 1.0** & -0.2 ± 1.1*  & 18.4 ± 2.4 & 30.5 ± 1.3 & -1.7 ± 0.8 & -1.4 ± 0.6 \\
\textbf{upper arm} & 32.8 ± 6.6 & 32.9 ± 1.7 & 0.1 ± 0.5**  & 0.3 ± 1.1**  & 17.6 ± 4.0 & 31.0 ± 1.1 & -1.6 ± 0.7 & -1.4 ± 0.5 \\
\textbf{forearm}   & 29.3 ± 3.6 & 34.1 ± 3.0 & -0.2 ± 0.9** & -0.2 ± 1.2** & 13.3 ± 4.9 & 30.0 ± 1.2 & -1.8 ± 0.7 & -1.5 ± 0.5 \\
\textbf{hip}       & 29.5 ± 4.4 & 29.0 ± 3.6 & -0.4 ± 1.2** & -0.2 ± 1.1** & 12.9 ± 2.5 & 24.0 ± 1.6 & -2.6 ± 0.7 & -2.1 ± 0.7 \\
\textbf{thigh}     & 31.0 ± 3.3 & 30.7 ± 2.4 & 0.1 ± 0.9**  & 0.4 ± 1.1**  & 19.3 ± 1.9 & 29.2 ± 2.2 & -1.9 ± 0.7 & -1.8 ± 0.6 \\
\textbf{calf}      & 31.7 ± 4.1 & 28.9 ± 3.2 & 0.0 ± 1.1**  & -0.1 ± 1.1** & 13.4 ± 1.8 & 26.9 ± 1.8 & -2.2 ± 0.7 & -1.8 ± 0.6 \\
\textbf{hand}      & -          & 27.0 ± 3.2 & -1.5 ± 0.9*  & -1.3 ± 0.9** & -          & 24.7 ± 4.0 & -2.7 ± 0.9 & -2.5 ± 0.4 \\
\textbf{foot}      & -          & 27.0 ± 4.9 & -1.5 ± 1.0*  & -1.3 ± 0.8** & -          & 24.7 ± 5.5 & -2.9 ± 0.7 & -2.5 ± 0.4 \\
\textbf{forehead}  & -          & 30.2 ± 2.5 & -0.5 ± 0.6** & -0.3 ± 0.5** & -          & 30.0 ± 2.7 & -1.4 ± 1.0 & -1.2 ± 0.9 \\
\textbf{neck}      & -          & -          & -0.6 ± 0.7** & -0.5 ± 0.8*  & -          & -          & -1.5 ± 1.0 & -1.2 ± 0.9 \\ \hline
\end{tabular}
\end{table}
\end{landscape}

\bibliographystyle{unsrturl}  
\bibliography{PCS_TJ2}  

\begin{thebibliography}{10}

\bibitem{chen_investigation_2020}
Xin Chen, Lixin Gao, Puning Xue, Jing Du, and Jing Liu.
\newblock Investigation of outdoor thermal sensation and comfort evaluation
  methods in severe cold area.
\newblock {\em Science of The Total Environment}, 749:141520, December 2020.
\newblock URL:
  \url{https://www.sciencedirect.com/science/article/pii/S004896972035049X},
  \href {https://doi.org/10.1016/j.scitotenv.2020.141520}
  {\path{doi:10.1016/j.scitotenv.2020.141520}}.

\bibitem{du_field_2020}
Jing Du, Cheng Sun, Qiuke Xiao, Xin Chen, and Jing Liu.
\newblock Field assessment of winter outdoor 3-{D} radiant environment and its
  impact on thermal comfort in a severely cold region.
\newblock {\em Science of The Total Environment}, 709:136175, March 2020.
\newblock URL:
  \url{https://www.sciencedirect.com/science/article/pii/S0048969719361716},
  \href {https://doi.org/10.1016/j.scitotenv.2019.136175}
  {\path{doi:10.1016/j.scitotenv.2019.136175}}.

\bibitem{fanger_thermal_1970}
P.~O. Fanger.
\newblock Thermal comfort. {Analysis} and applications in environmental
  engineering.
\newblock {\em Thermal comfort. Analysis and applications in environmental
  engineering.}, 1970.
\newblock URL: \url{https://www.cabdirect.org/cabdirect/abstract/19722700268}.

\bibitem{brager_evolving_2015}
Gail Brager, Hui Zhang, and Edward Arens.
\newblock Evolving opportunities for providing thermal comfort.
\newblock {\em Building Research \& Information}, 43(3):274--287, May 2015.
\newblock URL:
  \url{https://www.tandfonline.com/doi/full/10.1080/09613218.2015.993536},
  \href {https://doi.org/10.1080/09613218.2015.993536}
  {\path{doi:10.1080/09613218.2015.993536}}.

\bibitem{zhang_using_2015}
Hui Zhang, Edward Arens, Mallory Taub, Darryl Dickerhoff, Fred Bauman, Marc
  Fountain, Wilmer Pasut, David Fannon, Yongchao Zhai, and Margaret Pigman.
\newblock Using footwarmers in offices for thermal comfort and energy savings.
\newblock 104:233--243.
\newblock URL:
  \url{https://linkinghub.elsevier.com/retrieve/pii/S0378778815301067}, \href
  {https://doi.org/10.1016/j.enbuild.2015.06.086}
  {\path{doi:10.1016/j.enbuild.2015.06.086}}.

\bibitem{arens_are_2010}
Edward Arens, Michael~A. Humphreys, Richard~de Dear, and Hui Zhang.
\newblock Are ‘class {A}’ temperature requirements realistic or desirable?
\newblock {\em Building and Environment}, 45(1):4--10, 2010.
\newblock URL:
  \url{https://www.sciencedirect.com/science/article/pii/S036013230900078X},
  \href {https://doi.org/https://doi.org/10.1016/j.buildenv.2009.03.014}
  {\path{doi:https://doi.org/10.1016/j.buildenv.2009.03.014}}.

\bibitem{graham_lessons_2021}
Lindsay~t. Graham, Thomas parkinson, and Stefano schiavon.
\newblock Lessons learned from 20 years of {CBE}’s occupant surveys.
\newblock {\em Buildings and Cities}, 2(1):166--184, February 2021.
\newblock URL:
  \url{http://journal-buildingscities.org/articles/10.5334/bc.76/}, \href
  {https://doi.org/10.5334/bc.76} {\path{doi:10.5334/bc.76}}.

\bibitem{wang_individual_2018}
Zhe Wang, Richard de~Dear, Maohui Luo, Borong Lin, Yingdong He, Ali Ghahramani,
  and Yingxin Zhu.
\newblock Individual difference in thermal comfort: {A} literature review.
\newblock {\em Building and Environment}, 138:181--193, June 2018.
\newblock URL:
  \url{https://www.sciencedirect.com/science/article/pii/S0360132318302518},
  \href {https://doi.org/10.1016/j.buildenv.2018.04.040}
  {\path{doi:10.1016/j.buildenv.2018.04.040}}.

\bibitem{luo_high-density_2020}
Maohui Luo, Zhe Wang, Hui Zhang, Edward Arens, Davide Filingeri, Ling Jin, Ali
  Ghahramani, Wenhua Chen, Yingdong He, and Binghui Si.
\newblock High-density thermal sensitivity maps of the human body.
\newblock {\em Building and Environment}, 167:106435, January 2020.
\newblock URL:
  \url{https://linkinghub.elsevier.com/retrieve/pii/S0360132319306456}, \href
  {https://doi.org/10.1016/j.buildenv.2019.106435}
  {\path{doi:10.1016/j.buildenv.2019.106435}}.

\bibitem{yang_thermal_2022}
Bin Yang, Mengchun Wu, Zhe Li, Huangcheng Yao, and Faming Wang.
\newblock Thermal comfort and energy savings of personal comfort systems in low
  temperature office: {A} field study.
\newblock {\em Energy and Buildings}, 270:112276, September 2022.
\newblock URL:
  \url{https://linkinghub.elsevier.com/retrieve/pii/S0378778822004479}, \href
  {https://doi.org/10.1016/j.enbuild.2022.112276}
  {\path{doi:10.1016/j.enbuild.2022.112276}}.

\bibitem{chen_physiological_2019}
Xin Chen, Puning Xue, Lixin Gao, Jing Du, and Jing Liu.
\newblock Physiological and thermal response to real-life transient conditions
  during winter in severe cold area.
\newblock {\em Building and Environment}, 157:284--296, June 2019.
\newblock URL:
  \url{https://linkinghub.elsevier.com/retrieve/pii/S0360132319302367}, \href
  {https://doi.org/10.1016/j.buildenv.2019.04.004}
  {\path{doi:10.1016/j.buildenv.2019.04.004}}.

\bibitem{saedpanah_effects_2019}
Keivan Saedpanah, Mohsen Aliabadi, Majid Motamedzade, and Rostam Golmohammadi.
\newblock The effects of short-term and long-term exposure to extreme cold
  environment on the body’s physiological responses: {An} experimental study.
\newblock {\em Human Factors and Ergonomics in Manufacturing \& Service
  Industries}, 29(2):163--171, March 2019.
\newblock URL: \url{https://onlinelibrary.wiley.com/doi/10.1002/hfm.20770},
  \href {https://doi.org/10.1002/hfm.20770} {\path{doi:10.1002/hfm.20770}}.

\bibitem{wu_human_2021}
Jiansong Wu, Zhuqiang Hu, Zhaoxing Han, Yin Gu, Lin Yang, and Boyang Sun.
\newblock Human physiological responses of exposure to extremely cold
  environments.
\newblock {\em Journal of Thermal Biology}, 98:102933, May 2021.
\newblock URL:
  \url{https://linkinghub.elsevier.com/retrieve/pii/S0306456521001005}, \href
  {https://doi.org/10.1016/j.jtherbio.2021.102933}
  {\path{doi:10.1016/j.jtherbio.2021.102933}}.

\bibitem{thetkathuek_cold_2015}
Anamai Thetkathuek, Tanongsak Yingratanasuk, Wanlop Jaidee, and Wiwat
  Ekburanawat.
\newblock Cold {Exposure} and {Health} {Effects} {Among} {Frozen} {Food}
  {Processing} {Workers} in {Eastern} {Thailand}.
\newblock {\em Safety and Health at Work}, 6(1):56--61, March 2015.
\newblock URL:
  \url{https://linkinghub.elsevier.com/retrieve/pii/S2093791114000791}, \href
  {https://doi.org/10.1016/j.shaw.2014.10.004}
  {\path{doi:10.1016/j.shaw.2014.10.004}}.

\bibitem{zhu_cold_2022}
Yixiang Zhu, Ting Yang, Suijie Huang, Huichu Li, Jian Lei, Xiaowei Xue, Ya~Gao,
  Yixuan Jiang, Cong Liu, Haidong Kan, and Renjie Chen.
\newblock Cold temperature and sudden temperature drop as novel risk factors of
  asthma exacerbation: a longitudinal study in 18 {Chinese} cities.
\newblock {\em Science of The Total Environment}, 814:151959, March 2022.
\newblock URL:
  \url{https://linkinghub.elsevier.com/retrieve/pii/S0048969721070352}, \href
  {https://doi.org/10.1016/j.scitotenv.2021.151959}
  {\path{doi:10.1016/j.scitotenv.2021.151959}}.

\bibitem{tipton_human_2017}
M.~J. Tipton, N.~Collier, H.~Massey, J.~Corbett, and M.~Harper.
\newblock Human {Physiological} {Responses} to {Cold} {Exposure}.
\newblock {\em Experimental Physiology}, 102(11):1335--1355, November 2017.
\newblock URL: \url{http://doi.wiley.com/10.1113/EP086283}, \href
  {https://doi.org/10.1113/EP086283} {\path{doi:10.1113/EP086283}}.

\bibitem{wu_physiological_2021}
Jiansong Wu, Boyang Sun, Zhuqiang Hu, Letian Li, and Huizhong Zhu.
\newblock Physiological responses and thermal sensation during extremely cold
  exposure (-20 ℃).
\newblock {\em Building and Environment}, 206:108338, December 2021.
\newblock URL:
  \url{https://linkinghub.elsevier.com/retrieve/pii/S0360132321007356}, \href
  {https://doi.org/10.1016/j.buildenv.2021.108338}
  {\path{doi:10.1016/j.buildenv.2021.108338}}.

\bibitem{chen_underlying_2022}
Zhuangzhuang Chen, Peilin Liu, Xiaoshuang Xia, Lin Wang, and Xin Li.
\newblock The underlying mechanisms of cold exposure-induced ischemic stroke.
\newblock {\em Science of The Total Environment}, 834:155514, August 2022.
\newblock URL:
  \url{https://linkinghub.elsevier.com/retrieve/pii/S0048969722026109}, \href
  {https://doi.org/10.1016/j.scitotenv.2022.155514}
  {\path{doi:10.1016/j.scitotenv.2022.155514}}.

\bibitem{cai_cold_2016}
Jing Cai, Xia Meng, Cuicui Wang, Renjie Chen, Ji~Zhou, Xiaohui Xu, Sandie Ha,
  Zhuohui Zhao, and Haidong Kan.
\newblock The cold effects on circulatory inflammation, thrombosis and
  vasoconstriction in type 2 diabetic patients.
\newblock {\em Science of The Total Environment}, 568:271--277, October 2016.
\newblock URL:
  \url{https://linkinghub.elsevier.com/retrieve/pii/S0048969716312013}, \href
  {https://doi.org/10.1016/j.scitotenv.2016.06.030}
  {\path{doi:10.1016/j.scitotenv.2016.06.030}}.

\bibitem{wu_physiological_2021-1}
Jiansong Wu, Xinyu Ji, Zhuqiang Hu, Boyang Sun, and Letian Li.
\newblock Physiological responses and thermal sensation in the recovery period
  after extremely cold exposure.
\newblock {\em Building and Environment}, 200:107958, August 2021.
\newblock URL:
  \url{https://linkinghub.elsevier.com/retrieve/pii/S0360132321003620}, \href
  {https://doi.org/10.1016/j.buildenv.2021.107958}
  {\path{doi:10.1016/j.buildenv.2021.107958}}.

\bibitem{bandyopadhayaya_can_2020}
Shreetama Bandyopadhayaya, Rashmi Bundel, Shikhar Tyagi, Arvind Pandey, and
  Chandi~C. Mandal.
\newblock Can the aging influence cold environment mediated cancer risk in the
  {USA} female population?
\newblock {\em Journal of Thermal Biology}, 92:102676, August 2020.
\newblock URL:
  \url{https://linkinghub.elsevier.com/retrieve/pii/S0306456520304484}, \href
  {https://doi.org/10.1016/j.jtherbio.2020.102676}
  {\path{doi:10.1016/j.jtherbio.2020.102676}}.

\bibitem{wu_perceptual_2021}
Jiansong Wu, Lin Yang, Zhuqiang Hu, Fei Gao, and Xiaofeng Hu.
\newblock Perceptual response and cognitive performance during exposure to
  extremely cold environments.
\newblock {\em Energy and Buildings}, 251:111358, November 2021.
\newblock URL:
  \url{https://linkinghub.elsevier.com/retrieve/pii/S0378778821006423}, \href
  {https://doi.org/10.1016/j.enbuild.2021.111358}
  {\path{doi:10.1016/j.enbuild.2021.111358}}.

\bibitem{lichtenbelt_cold_2014}
Wouter van~Marken Lichtenbelt, Boris Kingma, Anouk van~der Lans, and Lisje
  Schellen.
\newblock Cold exposure – an approach to increasing energy expenditure in
  humans.
\newblock {\em Trends in Endocrinology \& Metabolism}, 25(4):165--167, April
  2014.
\newblock URL:
  \url{https://linkinghub.elsevier.com/retrieve/pii/S1043276014000101}, \href
  {https://doi.org/10.1016/j.tem.2014.01.001}
  {\path{doi:10.1016/j.tem.2014.01.001}}.

\bibitem{luo_effects_2022}
Wei Luo, Rick Kramer, Yvonne De~Kort, Pascal Rense, and Wouter Van
  Marken~Lichtenbelt.
\newblock effects of novel personal comfort systems on thermal comfort and
  thermophysiology.
\newblock {\em CLIMA 2022 conference}, page 2022: CLIMA 2022 The 14th REHVA
  HVAC World Congress, May 2022.
\newblock URL:
  \url{https://proceedings.open.tudelft.nl/clima2022/article/view/407}, \href
  {https://doi.org/10.34641/CLIMA.2022.407}
  {\path{doi:10.34641/CLIMA.2022.407}}.

\bibitem{nakamura_relative_2013}
Mayumi Nakamura, Tamae Yoda, Larry~I. Crawshaw, Momoko Kasuga, Yuki Uchida, Ken
  Tokizawa, Kei Nagashima, and Kazuyuki Kanosue.
\newblock Relative importance of different surface regions for thermal comfort
  in humans.
\newblock {\em European Journal of Applied Physiology}, 113(1):63--76, January
  2013.
\newblock URL: \url{http://link.springer.com/10.1007/s00421-012-2406-9}, \href
  {https://doi.org/10.1007/s00421-012-2406-9}
  {\path{doi:10.1007/s00421-012-2406-9}}.

\bibitem{zolfaghari_thermal_2010}
Alireza Zolfaghari and Mehdi Maerefat.
\newblock Thermal response of cutaneous thermoreceptors: {A} new criterion for
  the human body thermal sensation.
\newblock In {\em 2010 17th {Iranian} {Conference} of {Biomedical}
  {Engineering} ({ICBME})}, pages 1--4, Isfahan, Iran, November 2010. IEEE.
\newblock URL: \url{http://ieeexplore.ieee.org/document/5705004/}, \href
  {https://doi.org/10.1109/ICBME.2010.5705004}
  {\path{doi:10.1109/ICBME.2010.5705004}}.

\bibitem{castellani_human_2016}
John~W. Castellani and Andrew~J. Young.
\newblock Human physiological responses to cold exposure: {Acute} responses and
  acclimatization to prolonged exposure.
\newblock {\em Autonomic Neuroscience}, 196:63--74, April 2016.
\newblock URL:
  \url{https://linkinghub.elsevier.com/retrieve/pii/S1566070216300145}, \href
  {https://doi.org/10.1016/j.autneu.2016.02.009}
  {\path{doi:10.1016/j.autneu.2016.02.009}}.

\bibitem{deng_effects_2019}
Yue Deng, Bin Cao, Hecheng Yang, and Bin Liu.
\newblock Effects of local body heating on thermal comfort for audiences in
  open-air venues in 2022 {Winter} {Olympics}.
\newblock {\em Building and Environment}, 165:106363, November 2019.
\newblock URL:
  \url{https://linkinghub.elsevier.com/retrieve/pii/S0360132319305736}, \href
  {https://doi.org/10.1016/j.buildenv.2019.106363}
  {\path{doi:10.1016/j.buildenv.2019.106363}}.

\bibitem{wang_experimental_2020}
Haiying Wang, Manshu Xu, and Chunxiao Bian.
\newblock Experimental comparison of local direct heating to improve thermal
  comfort of workers.
\newblock {\em Building and Environment}, 177:106884, June 2020.
\newblock URL:
  \url{https://linkinghub.elsevier.com/retrieve/pii/S0360132320302432}, \href
  {https://doi.org/10.1016/j.buildenv.2020.106884}
  {\path{doi:10.1016/j.buildenv.2020.106884}}.

\bibitem{he_creating_2022}
Yingdong He, Thomas Parkinson, Edward Arens, Hui Zhang, Nianping Li, Jinqing
  Peng, John Elson, and Clay Maranville.
\newblock Creating alliesthesia in cool environments using personal comfort
  systems.
\newblock {\em Building and Environment}, 209:108642, February 2022.
\newblock URL:
  \url{https://linkinghub.elsevier.com/retrieve/pii/S0360132321010337}, \href
  {https://doi.org/10.1016/j.buildenv.2021.108642}
  {\path{doi:10.1016/j.buildenv.2021.108642}}.

\bibitem{he_meeting_2021}
Yingdong He, Nianping Li, Jiamin Lu, Na~Li, Qiaolin Deng, Chang Tan, and Jinbo
  Yan.
\newblock Meeting thermal needs of occupants in shared space with an adjustable
  thermostat and local heating in winter: {An} experimental study.
\newblock {\em Energy and Buildings}, 236:110776, April 2021.
\newblock URL:
  \url{https://linkinghub.elsevier.com/retrieve/pii/S0378778821000608}, \href
  {https://doi.org/10.1016/j.enbuild.2021.110776}
  {\path{doi:10.1016/j.enbuild.2021.110776}}.

\bibitem{luo_effectiveness_2022}
Wei Luo, Rick Kramer, Yvonne de~Kort, and Wouter van Marken~Lichtenbelt.
\newblock Effectiveness of personal comfort systems on whole-body thermal
  comfort – {A} systematic review on which body segments to target.
\newblock {\em Energy and Buildings}, 256:111766, February 2022.
\newblock URL:
  \url{https://linkinghub.elsevier.com/retrieve/pii/S0378778821010501}, \href
  {https://doi.org/10.1016/j.enbuild.2021.111766}
  {\path{doi:10.1016/j.enbuild.2021.111766}}.

\bibitem{luo_thermal_2018}
Maohui Luo, Edward Arens, Hui Zhang, Ali Ghahramani, and Zhe Wang.
\newblock Thermal comfort evaluated for combinations of energy-efficient
  personal heating and cooling devices.
\newblock {\em Building and Environment}, 143:206--216, October 2018.
\newblock URL:
  \url{https://linkinghub.elsevier.com/retrieve/pii/S0360132318304165}, \href
  {https://doi.org/10.1016/j.buildenv.2018.07.008}
  {\path{doi:10.1016/j.buildenv.2018.07.008}}.

\bibitem{tang_thermal_2022}
Yin Tang, Hang Yu, Kege Zhang, Kexin Niu, Huice Mao, and Maohui Luo.
\newblock Thermal comfort performance and energy-efficiency evaluation of six
  personal heating/cooling devices.
\newblock {\em Building and Environment}, 217:109069, June 2022.
\newblock URL:
  \url{https://linkinghub.elsevier.com/retrieve/pii/S0360132322003080}, \href
  {https://doi.org/10.1016/j.buildenv.2022.109069}
  {\path{doi:10.1016/j.buildenv.2022.109069}}.

\bibitem{yang_study_2020}
Hecheng Yang, Yue Deng, Bin Cao, and Yingxin Zhu.
\newblock Study on the local and overall thermal perceptions under nonuniform
  thermal exposure using a cooling chair.
\newblock {\em Building and Environment}, 176:106864, June 2020.
\newblock URL:
  \url{https://linkinghub.elsevier.com/retrieve/pii/S0360132320302237}, \href
  {https://doi.org/10.1016/j.buildenv.2020.106864}
  {\path{doi:10.1016/j.buildenv.2020.106864}}.

\bibitem{zhang_thermal_2010-2}
Hui Zhang, Edward Arens, Charlie Huizenga, and Taeyoung Han.
\newblock Thermal sensation and comfort models for non-uniform and transient
  environments: Part i: Local sensation of individual body parts.
\newblock 45(2):380--388.
\newblock URL:
  \url{https://linkinghub.elsevier.com/retrieve/pii/S0360132309001607}, \href
  {https://doi.org/10.1016/j.buildenv.2009.06.018}
  {\path{doi:10.1016/j.buildenv.2009.06.018}}.

\bibitem{zhang_thermal_2010-1}
Hui Zhang, Edward Arens, Charlie Huizenga, and Taeyoung Han.
\newblock Thermal sensation and comfort models for non-uniform and transient
  environments, part {II}: Local comfort of individual body parts.
\newblock 45(2):389--398.
\newblock URL:
  \url{https://linkinghub.elsevier.com/retrieve/pii/S0360132309001620}, \href
  {https://doi.org/10.1016/j.buildenv.2009.06.015}
  {\path{doi:10.1016/j.buildenv.2009.06.015}}.

\bibitem{zhang_thermal_2010}
Hui Zhang, Edward Arens, Charlie Huizenga, and Taeyoung Han.
\newblock Thermal sensation and comfort models for non-uniform and transient
  environments, part {III}: Whole-body sensation and comfort.
\newblock 45(2):399--410.
\newblock URL:
  \url{https://linkinghub.elsevier.com/retrieve/pii/S0360132309001619}, \href
  {https://doi.org/10.1016/j.buildenv.2009.06.020}
  {\path{doi:10.1016/j.buildenv.2009.06.020}}.

\bibitem{zhao_thermal_2014}
Yin Zhao, Hui Zhang, Edward~A. Arens, and Qianchuan Zhao.
\newblock Thermal sensation and comfort models for non-uniform and transient
  environments, part {IV}: Adaptive neutral setpoints and smoothed whole-body
  sensation model.
\newblock 72:300--308.
\newblock URL:
  \url{https://linkinghub.elsevier.com/retrieve/pii/S0360132313003181}, \href
  {https://doi.org/10.1016/j.buildenv.2013.11.004}
  {\path{doi:10.1016/j.buildenv.2013.11.004}}.

\bibitem{pasut_energy-efficient_2015}
Wilmer Pasut, Hui Zhang, Ed~Arens, and Yongchao Zhai.
\newblock Energy-efficient comfort with a heated/cooled chair: {Results} from
  human subject tests.
\newblock {\em Building and Environment}, 84:10--21, January 2015.
\newblock URL:
  \url{https://linkinghub.elsevier.com/retrieve/pii/S0360132314003473}, \href
  {https://doi.org/10.1016/j.buildenv.2014.10.026}
  {\path{doi:10.1016/j.buildenv.2014.10.026}}.

\bibitem{rissetto_personalized_2021}
Romina Rissetto, Marcel Schweiker, and Andreas Wagner.
\newblock Personalized ceiling fans: {Effects} of air motion, air direction and
  personal control on thermal comfort.
\newblock {\em Energy and Buildings}, 235:110721, March 2021.
\newblock URL:
  \url{https://linkinghub.elsevier.com/retrieve/pii/S0378778821000050}, \href
  {https://doi.org/10.1016/j.enbuild.2021.110721}
  {\path{doi:10.1016/j.enbuild.2021.110721}}.

\bibitem{arghand_individually_2022}
Taha Arghand, Arsen Melikov, Zhecho Bolashikov, Panu Mustakallio, and Risto
  Kosonen.
\newblock Individually controlled localized chilled beam with background
  radiant cooling system: {Human} subject testing.
\newblock {\em Building and Environment}, 218:109124, June 2022.
\newblock URL:
  \url{https://linkinghub.elsevier.com/retrieve/pii/S0360132322003614}, \href
  {https://doi.org/10.1016/j.buildenv.2022.109124}
  {\path{doi:10.1016/j.buildenv.2022.109124}}.

\bibitem{luo_can_2014}
Maohui Luo, Bin Cao, Xiang Zhou, Min Li, Jingsi Zhang, Qin Ouyang, and Yingxin
  Zhu.
\newblock Can personal control influence human thermal comfort? {A} field study
  in residential buildings in {China} in winter.
\newblock {\em Energy and Buildings}, 72:411--418, April 2014.
\newblock URL:
  \url{https://linkinghub.elsevier.com/retrieve/pii/S0378778814000061}, \href
  {https://doi.org/10.1016/j.enbuild.2013.12.057}
  {\path{doi:10.1016/j.enbuild.2013.12.057}}.

\bibitem{luo_underlying_2016}
Maohui Luo, Bin Cao, Wenjie Ji, Qin Ouyang, Borong Lin, and Yingxin Zhu.
\newblock The underlying linkage between personal control and thermal comfort:
  {Psychological} or physical effects?
\newblock {\em Energy and Buildings}, 111:56--63, January 2016.
\newblock URL:
  \url{https://linkinghub.elsevier.com/retrieve/pii/S0378778815303698}, \href
  {https://doi.org/10.1016/j.enbuild.2015.11.004}
  {\path{doi:10.1016/j.enbuild.2015.11.004}}.

\bibitem{parkinson_predicting_2021}
Thomas Parkinson, Hui Zhang, Ed~Arens, Yingdong He, Richard de~Dear, John
  Elson, Alex Parkinson, Clay Maranville, and Andrew Wang.
\newblock Predicting thermal pleasure experienced in dynamic environments from
  simulated cutaneous thermoreceptor activity.
\newblock {\em Indoor Air}, 31(6):2266--2280, November 2021.
\newblock URL: \url{https://onlinelibrary.wiley.com/doi/10.1111/ina.12859},
  \href {https://doi.org/10.1111/ina.12859} {\path{doi:10.1111/ina.12859}}.

\bibitem{brager_thermal_1998}
Gail~S. Brager and Richard~J. de~Dear.
\newblock Thermal adaptation in the built environment: a literature review.
\newblock {\em Energy and Buildings}, 27(1):83--96, February 1998.
\newblock URL:
  \url{https://linkinghub.elsevier.com/retrieve/pii/S0378778897000534}, \href
  {https://doi.org/10.1016/S0378-7788(97)00053-4}
  {\path{doi:10.1016/S0378-7788(97)00053-4}}.

\bibitem{zhang_review_2015}
Hui Zhang, Edward Arens, and Yongchao Zhai.
\newblock A review of the corrective power of personal comfort systems in
  non-neutral ambient environments.
\newblock {\em Building and Environment}, 91:15--41, September 2015.
\newblock URL:
  \url{https://linkinghub.elsevier.com/retrieve/pii/S0360132315001225}, \href
  {https://doi.org/10.1016/j.buildenv.2015.03.013}
  {\path{doi:10.1016/j.buildenv.2015.03.013}}.

\bibitem{deng_effects_2020}
Yue Deng, Bin Cao, Bin Liu, and Yingxin Zhu.
\newblock Effects of local heating on thermal comfort of standing people in
  extremely cold environments.
\newblock {\em Building and Environment}, 185:107256, November 2020.
\newblock URL:
  \url{https://linkinghub.elsevier.com/retrieve/pii/S0360132320306272}, \href
  {https://doi.org/10.1016/j.buildenv.2020.107256}
  {\path{doi:10.1016/j.buildenv.2020.107256}}.

\bibitem{ju_development_2020}
Yi~Ju, Bin Cao, and Xinyuan Ju.
\newblock Development of a similarity function to evaluate interpersonal
  differences in body thermal sensitivity distribution patterns.
\newblock Seoul, South Korea, 2020.

\bibitem{zhang_effect_2007}
Yufeng Zhang and Rongyi Zhao.
\newblock Effect of local exposure on human responses.
\newblock {\em Building and Environment}, 42(7):2737--2745, July 2007.
\newblock URL:
  \url{https://linkinghub.elsevier.com/retrieve/pii/S0360132306001946}, \href
  {https://doi.org/10.1016/j.buildenv.2006.07.014}
  {\path{doi:10.1016/j.buildenv.2006.07.014}}.

\bibitem{kaikaew_sex_2018}
Kasiphak Kaikaew, Johanna~C. van~den Beukel, Sebastian~J.C.M.M. Neggers,
  Axel~P.N. Themmen, Jenny~A. Visser, and Aldo Grefhorst.
\newblock Sex difference in cold perception and shivering onset upon gradual
  cold exposure.
\newblock {\em Journal of Thermal Biology}, 77:137--144, October 2018.
\newblock URL:
  \url{https://linkinghub.elsevier.com/retrieve/pii/S0306456518301918}, \href
  {https://doi.org/10.1016/j.jtherbio.2018.08.016}
  {\path{doi:10.1016/j.jtherbio.2018.08.016}}.

\bibitem{maeda_seasonal_2005}
Takafumi Maeda, Toshio Kobayashi, Kazuko Tanaka, Akihiko Sato, Shin-Ya Kaneko,
  and Masatoshi Tanaka.
\newblock Seasonal differences in physiological and psychological responses to
  hot and cold environments in the elderly and young males.
\newblock In {\em Elsevier {Ergonomics} {Book} {Series}}, volume~3, pages
  35--41. Elsevier, 2005.
\newblock URL:
  \url{https://linkinghub.elsevier.com/retrieve/pii/S1572347X05800072}.

\bibitem{luo_indoor_2016}
Maohui Luo, Wenjie Ji, Bin Cao, Qin Ouyang, and Yingxin Zhu.
\newblock Indoor climate and thermal physiological adaptation: {Evidences} from
  migrants with different cold indoor exposures.
\newblock {\em Building and Environment}, 98:30--38, March 2016.
\newblock URL:
  \url{https://linkinghub.elsevier.com/retrieve/pii/S0360132315302134}, \href
  {https://doi.org/10.1016/j.buildenv.2015.12.015}
  {\path{doi:10.1016/j.buildenv.2015.12.015}}.

\bibitem{cao_too_2016}
Bin Cao, Maohui Luo, Min Li, and Yingxin Zhu.
\newblock Too cold or too warm? {A} winter thermal comfort study in different
  climate zones in {China}.
\newblock {\em Energy and Buildings}, 133:469--477, December 2016.
\newblock URL:
  \url{https://linkinghub.elsevier.com/retrieve/pii/S0378778816308969}, \href
  {https://doi.org/10.1016/j.enbuild.2016.09.050}
  {\path{doi:10.1016/j.enbuild.2016.09.050}}.

\bibitem{LI2023109798}
Sishi Li, Xinyu Jia, Bin Cao, Bin Liu, and Yingxin Zhu.
\newblock Thermal comfort characteristics and heating demand of people with
  different activity status during extremely cold exposure.
\newblock {\em Building and Environment}, 228:109798, 2023.
\newblock URL:
  \url{https://www.sciencedirect.com/science/article/pii/S0360132322010289},
  \href {https://doi.org/https://doi.org/10.1016/j.buildenv.2022.109798}
  {\path{doi:https://doi.org/10.1016/j.buildenv.2022.109798}}.

\end{thebibliography}

\end{document}